\documentclass[ parskip = half, twoside, footinclude = false]{scrartcl}
\pdfoutput=1 
\usepackage{ifthen} 

\newboolean{boldeqnitalic}
\setboolean{boldeqnitalic}{true}
\usepackage{stylefiles/defdb1} 

\usepackage{stylefiles/rublkm_paper} 
\usepackage{caption}
\usepackage{newfloat}
\usepackage{subcaption}
\usepackage{xcolor}
\usepackage{doi}
\usepackage{enumitem}

\usepackage{tikz}
\usepackage{pgfplots}
\usepackage{pgfplotstable}
\usepgfplotslibrary{groupplots}
\usetikzlibrary{shapes,arrows.meta}
\usetikzlibrary{external}
\usepackage{algorithm,algorithmic}
\usetikzlibrary{calc}

\newboolean{genfig}
\setboolean{genfig}{false}
\ifthenelse{\boolean{genfig}}
{\usepackage{poisson}}
{}
\newcommand{\logopath}{figures_logo} 

\DeclareFloatingEnvironment[
  fileext=lob,
  listname={List of Boxes},
  name=Box,
  placement=htp,
]{BOX}

\begin{document}


\thispagestyle{empty}


\begin{figure}[htb]
\unitlength1cm
\begin{picture}(21,1.25)

\put(-2.0, 2.45){\textcolor{\headercolor}{\sffamily\textbf{Chair of Continuum Mechanics}}}
\put(-2.0, 1.90){\textcolor{\headercolor}{\sffamily{Prof.~Dr.-Ing.~habil. Daniel Balzani}}}
\put(12.0, 1.6){\includegraphics{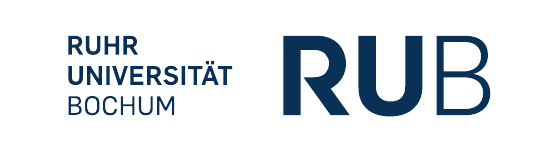}}
\end{picture}
\end{figure}


\vspace{3cm}

{\Huge
Evolving Microstructures in Relaxed Continuum Damage Mechanics for Strain Softening\\
}
[PREPRINT]

\vspace{5mm}

Maximilian Köhler, Daniel Balzani

\vfill


\clearpage
\setcounter{page}{1}
\begin{center}
{\LARGE Evolving Microstructures in Relaxed Continuum Damage Mechanics for the Modeling of Strain Softening}

\vspace{5mm}

Maximilian Köhler$^{1}$, Daniel Balzani$^{1\star}$

\vspace{3mm}

{\small $^1$Chair of Continuum Mechanics, Ruhr-Universit\"at Bochum,\\ 
Universit\"atsstra{\ss}e~150, 44801~Bochum, Germany}\\[3mm]

\vspace{3mm}

{\small ${}^{\star}$E-mail address of corresponding author: daniel.balzani@rub.de}

\vspace{10mm}

\begin{minipage}{15.0cm}
\textbf{Abstract}\hspace{3mm}
A new relaxation approach is proposed which allows for the description of stress- and strain-softening at finite strains. 
The model is based on the construction of a convex hull replacing the originally non-convex incremental stress potential which in turn represents damage in terms of the classical $(1-D)$ approach. 
This convex hull is given as the linear convex combination of weakly and strongly damaged phases and thus, it represents the homogenization of a microstructure bifurcated in the two phases. 
As a result thereof, damage evolves in the convexified regime mainly by an increasing volume fraction of the strongly damaged phase. 
In contrast to previous relaxed incremental formulations in \citet{GurMie:2011:edm} and \citet{BalOrt:2012:riv}, where the convex hull has been kept fixated after construction, here, the strongly damaged phase is allowed to elastically unload upon further loading. 
At the same time, its volume fraction increases nonlinearly within the convexified regime. 
Thus, strain-softening in the sense of a decreasing stress with increasing strain can be modeled. 
The major advantage of the proposed approach is that it ensures mesh-independent structural simulations without the requirement of additional length-scale related parameters or nonlocal quantities, which simplifies an implementation using classical material subroutine interfaces. 
In this paper, focus is on the relaxation of one-dimensional models for fiber damage which are combined with a microsphere approach to allow for the description of three-dimensional fiber dispersions appearing in fibrous materials such as soft biological tissues. 
Several numerical examples are analyzed to show the overall response of the model and the mesh-independence of resulting structural calculations. 
%

\end{minipage}
\end{center}

\medskip{}

\section{Introduction}
The phenomenological, macroscopic description of microscopic damage requires two primary effects to be modeled, namely, stress-softening and strain-softening.
The prior, also known as the Mullin's effect, leads to a reduced stiffness with increasing loads and thus, a lowered stress-strain curve.
If this reduced stiffness reaches an intensity where stresses start to decrease with increasing strains, then the response is referred to as strain-softening.
The challenging task is to formulate models that can capture both effects, as well as providing a well posed behavior. 
Often, models agreeing well with experiments can be formulated, 
however, obtaining a well posed behavior is rather difficult. 
Especially the strain--softening case possesses problematic mathematical structures if an incremental variational framework is used. 
This framework presented by a series of papers \cite{Hac:1997:gsm,OrtSta:1999:vfv,OrtRep:1999:nem,CarHacMie:2002:ncp,MieTheLev:2002:vfr} considers formulations for dissipative materials and 
constructs thermodynamically consistent potentials, which are pseudo-elastic in each incremental step.
However, if the underlying generalized energy density of the potential is non-convex, which is needed in order to describe strain-softening in a single increment, the solution of discretized boundary value problems becomes dependent on the discretization. 
In case of finite element discretization, the solution becomes dependent on the number and arrangement of the finite elements which is referred to as mesh-dependency.
Different techniques exists to overcome this mesh dependency. 
Gradient enhanced formulations and non-local material models have been proposed to provide
a remedy to the ill posed behavior, see e.g. \cite{Wu:2017:upt,WafPolMenBla:2014:glc,JunRieBal:2022:ern,PeeGeedeBre:2001:ccn} and references therein.
These non-local formulations introduce a length-scale parameter which is hard to identify and which has numerical as well as physical constraints. 
The direct description of the micromechanical effects that describe 
the damage behavior at the macroscale shows similar
problems. 
As stated in~\citet{GitAskSlu:2007:rve}, there exists no representative volume element to describe strain-softening. 
Other publications, e.g.,~\citet{LiaRenLi:2018:mmq}, where a scaled damage law has been used to overcome the problem, observed the same. 

Recently, a rather new modeling approach has been presented in~\citet{GorSchHryBoh:2021:cad}, that yields a priori a convex formulation. 
There, damage has been described by a compliance tensor, however, the approach is so far restricted 
to a pure stress-softening description, which is referred to as hardening in the original publication. 
Another approach in handling the non-convexity and thus, the mesh dependency of the formulation, is to use relaxation, which will be the focus of this paper. 
Within the relaxation framework an originally non-convex formulation is replaced by a (semi)convex envelope. 
Here, (semi)convex envelope denotes the largest (semi)convex function below the original function. 
The attractiveness of this regularization approach lies in the fact that no additional parameters are introduced to the model. 
Another advantageous property of relaxation is that in specific cases homogenized microstructures are described by the (semi)convex envelope. 
In~\citet{SchJunHac:2020:vrd} the procedure has been flipped, meaning, that the microstructures are described (emulated) and based on them, a (semi)convex hull is formulated. 
The desirable properties of relaxation theory lead to applications in the continuum mechanical context, see e.g. \citet{LamMieDet:2003:ern} for application in small strain plasticity and \citet{BarCarHacHop:2004:erm} for application in large strain plasticity, where the classical route is pursued, i.e. formulation of the (semi)convex hull and homogenized microstructures as a byproduct. 
Within continuum damage mechanics, \citet{GurMie:2011:edm} used relaxation for the small strain setting and \citet{BalOrt:2012:riv} in the finite strain setting. 
However, relaxation needs the (semi)convex envelope of the non-convex potential and this can be a sophisticated task to solve, especially in the finite strain setting. 
Due to certain simplifications, \cite{GurMie:2011:edm} reported that the convex hull can be computed offline, i.e. before the finite element simulation starts.
In contrast to this, \cite{BalOrt:2012:riv} solved a multi-modal optimization problem with a multistart Newton, leading to a rather expensive material law formulation with a strong sensitivity with respect to the obtained minimum. 
In \citet{SchBal:2016:riv} the multistart strategy was enhanced by choosing the multistart points based on an evolution strategy. 
Since both contributions can not guarantee to find the global minimum, a new, adaptive, discrete approach has been presented in \citet{KohNeuPetPetBal:2021:easa} to construct the convex hull, which is based on a variation of Graham's scan \cite[Sec 33.3]{CorLeiRivSte:2001:ita}.
Within the variational context it was written down in \citet[Remark 9.9 (ii)]{Bar:2015:nmn}
The new procedure guarantees to find the convex hull of a discretized one-dimensional potential and is orders of magnitudes faster.
Note that the discretized convex envelope only corresponds in the limit to the analytic convex envelope.
It appears that the approaches presented in~\citet{SchJunHac:2020:vrd} and~\citet{KohNeuPetPetBal:2021:easa} are the first relaxed formulations to allow for the description of strain-softening. 
Whereas the first is based on significant simplifications at the microstructural level, the second incorporates fiber failure as the driving mechanism for strain-softening. 
Resulting from the associated discontinuities, quite expensive numerical quadrature is required in the microsphere approach proposed there to obtain a suitable macroscopic response.

In the approach proposed in this paper, we follow a different strategy. 
Instead of fixating the convex hull for all following incremental steps, here, we allow for an updated convexification. 
This repeated convexification is made feasible by the computational savings of the algorithmic approach in~\citet{KohNeuPetPetBal:2021:easa}. 
As will be shown, the strongly damaged phase elastically unloads upon damage evolution and a nonlinear evolution of the phase fraction of the strongly damaged phase is observed. 
This leads to the desired strain-softening at the macroscale while ensuring convexity in the incremental problems and thus, rendering finite element calculations mesh-independent. 

This contribution is structured as follows: In section \ref{sec:math} the incremental variational framework of continuum damage mechanics and its relaxation from \cite{BalOrt:2012:riv} is briefly recapitulated.
The next section~\ref{sec:reconvexify} covers the novel model which considers relaxation in each incremental step in order to describe strain-softening.
In section \ref{sec:app}, applications and examples are presented in order to show the mesh independence of the new model. 
Here, the model is tested in the one-dimensional, two-dimensional plane strain, and three-dimensional case, where the higher-dimensional response is obtained by integrating the one-dimensional material law over the unit sphere. 
Section \ref{sec:conclusion} concludes the paper and states potential future topics.

\section{Mathematical Framework for Damage}
\label{sec:math}

The physical body of interest $\mathcal{B}$ is parameterized in $\bX$, which corresponds to the reference configuration of the body.
In each time instance $t \in \mathbb{R}^+$, a deformation $\boldsymbol{\varphi}_t : \mathcal{B} \rightarrow \mathcal{B}_t$ maps the reference to the actual (current, deformed) configuration of the body $\mathcal{B}_t \subset \mathbb{R}^3$.
Here, the actual configuration is parameterized in $\bx \in \mathcal{B}_t$.
The deformation gradient $\bF$ defines important kinematic quantities and is defined by $
\bF(\bX) = \text{Grad} \boldsymbol{\varphi}_t(\bX)
$. 
The material frame indifference is an essential continuum mechanical principle that can be fulfilled automatically if strain energy densities $\psi$ depend on the right Cauchy-Green tensor~$\bC$, i.e.
\eb
\psi \defeq \psi(\bC) \qquad \text{with} \quad \bC=\bF^T \bF.
\ee
The strain energy density is often formulated in terms of the invariants of~$\bC$, i.e. 
$I_1 = \text{tr} \bC$, $I_2 = \text{tr}[\text{Cof} \bC]$, $I_3 = \det \bC$, 
such that $\psi\defeq \psi\left(I_1,I_2,I_3\right)$. 
Within the scalar valued phenomenological approach of continuum damage mechanics, the strain energy density is defined as 
\eb
\psi(\bC,\beta) = (1-D(\beta))\psi^0(\bC),
\ee
yielding the so-called $(1-D)$ approach, where $D$ is a scalar-valued damage function. 
This model goes back to \citet{Kac:1958:rtc}, see \citet[section 2.1.1]{Mur:2012:cdm} for historic details. 
The damage function $D(\beta)$ 
depends on an internal variable~$\beta$ describing damage-related microstructure changes and is considered to be a quantity defined in terms of the thermodynamic force. 
Evaluating
the reduced Clausius-Duhem inequality for isothermal processes and application of the standard argument of rational continuum mechanics leads to the dissipation potential 
\eb
\label{eq:dissipation-potential}
\phi \defeq \psi^0 \dot{D} \geq 0
\ee
The discontinuous damage approach in the sense of \citet{Mie:1995:dcd} is considered here where the internal variable and the damage function are
%
\eb
\label{eq:damagefunction}
\beta \defeq \max_{s\leq t} \left[\psi^0(\bC)\right]
\quad\mbox{and}\quad
D(\beta) = D_{\infty} \left[1-\exp\left(\frac{-\beta}{D_0}\right)\right], 
\ee
%
respectively. 
Here, $D_{\infty}$ and $D_0$ denote the maximum possible damage and damage saturation
parameter, respectively, which are material dependent parameters.

\subsection{Incremental Formulation}

An incremental variational formulation for finite strain
continuum damage mechanics has been derived and discussed in detail in \cite{BalOrt:2012:riv}, which is now briefly recapitulated. 
In order to obtain a potential which is pseudo-elastic for discrete time increments, 
the incremental variational framework is employed. 
Hence, the incremental stress potential 
%
\eb
W(\bF_{k+1}) = \inf_{\beta_{k+1}}\left[\mathcal{W}(\bF_{k+1},\beta_{k+1})\right] 
\quad \text{where} \quad 
\mathcal{W}(\bF_{k+1},\beta_{k+1}) = \int_{t_k}^{t_{k+1}} \dot{\psi} + \phi \ \text{d}t
\ee
is considered such that the internal variable is updated
via
$
\beta_{k+1} = \arg \min \mathcal{W}.
$
In the case where damage evolves, the internal variable results in~$\beta_{k+1} = \psi^0(\bF_{k+1})$ and it remains constant otherwise.
Following \cite{BalOrt:2012:riv}, the incremental stress potential $W(\bF_{k+1})$, also often referred to as condensed or reduced energy, can be analytically derived 
as follows
\eb
\label{eq:isp}
W(\bF) = \psi(\bF,D) - \psi(\bF_k,D_k) + \beta D - \beta_k D_k - \overline{D} + \overline{D}_k.
\ee
The idea of pointwise minimization that yields a per incremental step pseudo-elastic potential goes back to a series of paper, see e.g.,  \cite{Hac:1997:gsm,OrtSta:1999:vfv,OrtRep:1999:nem,CarHacMie:2002:ncp,MieTheLev:2002:vfr}. 
Here, the abbreviations $\bF \defeq \bF_{k+1}, D\defeq D(\beta_{k+1}), D_k \defeq D(\beta_k), \overline{D} \defeq \overline{D}(\beta_{k+1}),$ and $\overline{D}_k \defeq \overline{D}(\beta_{k})$ are introduced, where $\overline{(\cdot)}$ denotes the anti derivative. 
Note that the subscript $(\cdot)_{k+1}$ is skipped in the following. 
Now, the principle of minimum potential energy can be employed
\eb
\inf_{\boldsymbol{\varphi}} \left[\Pi(\boldsymbol{\varphi})| \boldsymbol{\varphi}=\hat{\boldsymbol{\varphi}} \text{ on } \partial\mathcal{B}_{\varphi}, \boldsymbol{t}=\hat{\boldsymbol{t}} \text{ on } \partial\mathcal{B}_{\sigma}\right]
\quad\mbox{with}\quad
\Pi(\boldsymbol{\varphi}) \defeq \int_{\mathcal{B}} W(\bF(\boldsymbol{\varphi})) \ \text{d}V + \Pi^{\text{ext}}(\boldsymbol{t}(\boldsymbol{\varphi}))
\ee
to obtain the displacement field $\bu = \bx - \bX$. 
This minimization principle can be recast by a vanishing first variation which yields 
in the total Lagrangian setting
\eb
\label{eq:weakform}
\int_{\mathcal{B}} \bP : \text{Grad} \delta \bu \text{ d}V - \int_{\partial\mathcal{B}_{\sigma}} \hat{\bt} \cdot \delta \bu \text{ d}S = 0. 
\ee
%
Herein, the first Piola-Kirchhoff stresses~$\bP$ and the associated tangent moduli~$\IA$ are
\eb
\bP = \frac{\partial W(\bF)}{\partial \bF} \qquad \mathbb{A} = \frac{\partial \bP}{\partial \bF} = \frac{\partial^2 W(\bF)}{\partial \bF \partial \bF}.
\ee
The physical Cauchy stresses can be computed by~$\Bsigma=J^{-1}\bP\bF^{\text{T}}$. 
Discretizing the problem with finite elements leads to an ill-posed behavior due to the properties of $W(\bF)$. 
The incremental stress potential lacks (semi)convexity properties and thus, multiple stationary points exist. 
In order to obtain a well-posed response, the problem is now relaxed.

\subsection{Relaxation}

Relaxation replaces the non-convex potential $W(\bF)$ by its (semi)convex envelope.
In \cite{GurMie:2011:edm} and \cite{BalOrt:2012:riv} relaxed incremental variational formulations for damage were derived in the small and finite strain setting, respectively. 
Within both publications, the one-dimensional relaxation was examined as well as in this contribution. 
Thus, $W(\bF)$ is now restricted to the one-dimensional case $W(F)$, where all semiconvex notions (quasiconvexity, rank-one convexity, polyconvexity) coincide with convexity. 
A function $W(F)$ is said to be convex if 
\eb
\label{eq:convexity}
W(\xi F^+ + (1-\xi)F^-) \leq \xi W(F^+) + (1-\xi) W(F^-) \quad \text{with } \quad
\xi \in (0,1) \text{ and } F^+ \neq F^-,
\ee
which can be expressed in terms of the first derivative, if $W(F)$ is sufficiently regular (which is the case for continuum damage mechanics), i.e., 
\eb
\label{eq:convexification-tangent}
\left(W'(F^+) - W'(F^-)\right)(F^+-F^-) \geq 0.
\ee
Here, the points $F^+$ and $F^-$ denote two arbitrary points whose linear connection must lie above or on the function values of $W(F^*)$ at $F^*=\xi F^+ + (1-\xi)F^-$.
The convex envelope $W^C(F)$ of $W(F)$ is now defined as the largest convex function below $W(F)$.
This can be mathematically expressed as
\eb
\label{eq:convex-optimization}
W^C(F)=\inf_{\xi,d}\left[\overline{W}(F,d,\xi)\right] 
\quad\mbox{with}\quad 
\overline{W}= \xi W(F^+) + (1-\xi)W(F^-), 
\ee
where $F^+$ and $F^-$ are the supporting points of the convexified regime. 
They correspond to the lowest linear connection of two function values of $W(F)$, i.e., 
\eb
\label{eq:convexcombinationF}
F \defeq \xi F^+ + (1-\xi) F^-, 
\quad\mbox{where}\quad
\left\{\begin{aligned}
F^+ &\defeq F(1+(1-\xi)d), \\
F^- &\defeq F(1-\xi d).
\end{aligned}
\right.
\ee
The parameter $\xi$ denotes a volume fraction that states how much a given $F$ attains the value $F^+$ and $F^-$, respectively.
Further, the parameter $d$ corresponds mathematically to a distance between $F^+$ and $F^-$ and physically to a microscopic bifurcation intensity.
The points $F^+$ and $F^-$ turn out to be minimizers of the original problem between which the given deformation gradient $F$ tends to oscillate.
These points describe microstructures consisting of a weakly and strongly damaged phase with associated internal variables $\beta^+$ and $\beta^-$, respectively. 
This means that instead of a continuous evolution of damage throughout the material, as described by the original model, the relaxed model considers the nucleation of a microstructure when reaching~$F^-$ and an increase of the fraction of the strongly damaged phase upon damage evolution.
For in-depth mathematical details about the description of microstructures within relaxation, we refer the interested reader to \cite[Chapter 4]{Ped:2000:vmn},\cite[Chapter 9]{Bar:2015:nmn},\cite[Section 6]{KohStr:1986:odra}.

\section{Evolving Relaxed Formulation for Damage 
\label{sec:reconvexify}}

To motivate the approach proposed here, it is reasonable to reconsider the relaxed model and its microscopic interpretation in more detail. 
In fact, the convexification described above is valid for each incremental step separately, since $W(F)$~changes in each increment. 
In general, this happens due to the minimization (condensation, reduction) of the generalized energy density $\mathcal{W}(F,\beta)$ w.r.t. $\beta$, which is performed in each incremental step. 
So far, in the context of damage, the weakly and strongly damaged phases, which were identified whenever a loss of convexity is found first, were kept fixated. 
This means that the deformation states~$F^-$ and~$F^+$ as well as the associated damage states were frozen for subsequent incremental steps. 
The according microscopic interpretation is that damage evolution happens purely as a result from the increasing volume fraction of the strongly damaged phase, not an evolution of the states of the phases ifself. 
This procedure is convenient as the convexification has only to be performed until the loss of convexity is reached.
However, due to the simple linear convex combination of the fixated phase states, strain softening can not be described. 
This has been a major shortcoming of previous relaxation schemes for damage. 
Therefore, it is reasonable to allow for some evolution within the phase states themselves and to reconvexify the energy in subsequent incremental steps. 
This will be the core idea of the approach presented here. 
In fact, it has been an open research question for quite some time how to proceed with subsequent incremental steps. 
A series of paper raised the question how to formulate the subsequent time steps, e.g. \citet[section 10.4]{BarCarHacHop:2004:erm}.
Mathematical details about this can be found in \citet{Mie:2003:eri} and mechanical applications in \citet{HacHeiMie:2012:mel} and \citet{KocHac:2011:elfa}.
The aforementioned contributions share the thought, that the dissipation potential is reformulated in terms of a metric between two probability distributions.
The metric measures the transport from one statistical measure to another, where the measure is a byproduct of relaxation.
Within relaxation methods, the points $F^+$ and $F^-$ form the two point masses of Dirac delta distributions $\delta_{F^-}, \delta_{F^+}$, such that
\eb
\Xi = \xi \delta_{F^+} + (1-\xi) \delta_{F^-} 
\quad\mbox{and}\quad 
\langle \Xi, g \rangle = \xi g\left(F^+\right) + (1-\xi) g\left(F^-\right),
\ee
where $\langle \Xi, \cdot \rangle$ is the action of $\Xi$ on a continuous function $g$.
Here, $\Xi$ is a gradient Young measure (GYM), denoted as $\Xi \in \text{GYM}$, fulfilling the following properties \cite[Page 268]{Bar:2015:nmn}
\eb
\label{eq:GYM}
\langle \Xi, 1\rangle = 1, \quad
\langle \Xi, \text{Id} \rangle = F 
\quad\mbox{and,}\quad 
\langle \Xi, g \rangle = \frac{1}{\text{Vol}\left(\mathcal{B}_{\text{rep}}\right)}\int_{\mathcal{B}_{\text{rep}}} g(F) \ \text{d}V,
\ee
where $\text{Id}$ and $\mathcal{B}_{\text{rep}}$ denote the identity mapping and a representative microscopic subdomain of the body, respectively. 
Now, according to \cite{Mie:2003:eri, HacHeiMie:2012:mel, KocHac:2011:elfa}, the dissipation potential~$\phi \defeq \phi\left(\beta,\dot{\beta}\right)$ is replaced by 
\eb
\phi \defeq \phi^\star\left(\xi, \dot{\xi}, \beta^{\pm}, \dot{\beta^{\pm}}\right),
\ee
such that the transition from~$\Xi_k$ to~$\Xi$ is minimal.
As stated in \cite{HacHeiMie:2012:mel}, \citet{FraGar:2006:vvpa} used the Wasserstein distance for this purpose within a relaxed damage formulation, see section 5.2 in the mentioned publication.

\begin{figure}[t]
    \centering
    \ifthenelse{\boolean{genfig}}
    {\tikzsetfigurename{energy-neohooke}
\begin{tikzpicture}
    \begin{axis}[name=sigmaF,width={0.8\textwidth}, height={6cm},
    ylabel={$W(F)$},
    xlabel={$F$},
    axis lines=left,
    xtick={5,10,15,20},
    ytick={0,0.5,1,1.5},
    ymax=1.8,
    xmax=24,
    ymin=-0.5,
    axis lines = middle,
    axis line style=-{Latex[scale=1.5]},
    ]
        \addplot+[no markers, color={red!10!gray}, thick]
            table{figures/tikz/rawdata/energy_neohooke_lambda=0.0_mu=0.5_Dinfty=0.99_D0=0.5/F_W_1.08.txt}
            node[red!10!gray, pos = 1,yshift=0.0cm,xshift=0.8cm] {\smaller $\beta=0.003$};
        \addplot+[no markers, color={red!20!gray}, thick]
            table{figures/tikz/rawdata/energy_neohooke_lambda=0.0_mu=0.5_Dinfty=0.99_D0=0.5/F_W_1.6.txt}
            node[red!20!gray, pos = 1,yshift=0.0cm,xshift=0.5cm] {\smaller $0.155$};
        \addplot+[no markers, color={red!50!gray}, thick]
            table{figures/tikz/rawdata/energy_neohooke_lambda=0.0_mu=0.5_Dinfty=0.99_D0=0.5/F_W_2.2.txt}
            node[red!50!gray, pos = 1,yshift=0.0cm,xshift=0.5cm] {\smaller $0.566$};
        \addplot+[no markers, color={red}, thick]
            table{figures/tikz/rawdata/energy_neohooke_lambda=0.0_mu=0.5_Dinfty=0.99_D0=0.5/F_W_2.8.txt}
            node[red, pos = 1,yshift=0.0cm,yshift=-0.1cm,xshift=0.5cm] {\smaller $1.195$};
        \addplot+[no markers, color={red!10!gray}, thin,dashed]
            table{figures/tikz/rawdata/energy_neohooke_lambda=0.0_mu=0.5_Dinfty=0.99_D0=0.5/Fplusminus_Wplusminus_1.08.txt}
            ;
        \addplot+[no markers, color={red!20!gray}, thin, dashed]
            table{figures/tikz/rawdata/energy_neohooke_lambda=0.0_mu=0.5_Dinfty=0.99_D0=0.5/Fplusminus_Wplusminus_1.6.txt}
            ;
        \addplot+[no markers, color={red!50!gray}, thin, dashed]
            table{figures/tikz/rawdata/energy_neohooke_lambda=0.0_mu=0.5_Dinfty=0.99_D0=0.5/Fplusminus_Wplusminus_2.2.txt}
            ;
        \addplot+[no markers, color={red}, thin, dashed]
            table{figures/tikz/rawdata/energy_neohooke_lambda=0.0_mu=0.5_Dinfty=0.99_D0=0.5/Fplusminus_Wplusminus_2.8.txt}
            ;
    \end{axis}
\end{tikzpicture}}
    {\includegraphics{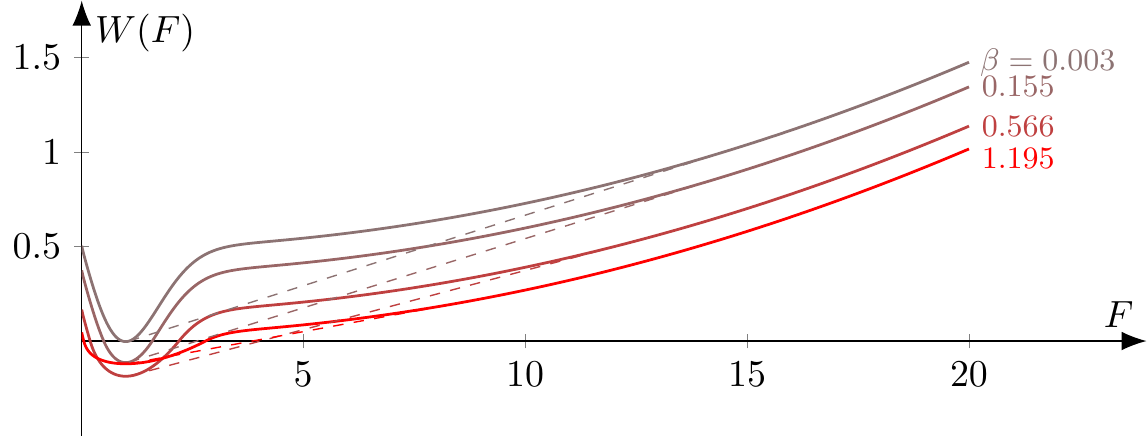}}
    \caption{Development of the incremental stress potential $W(F)$ vs. internal variable~$\beta$ for different point-wise condensed states.}
    \label{fig:energy_neohooke}
\end{figure}
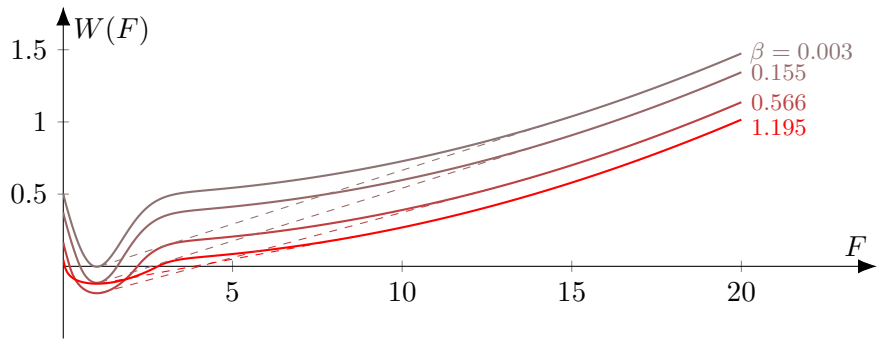

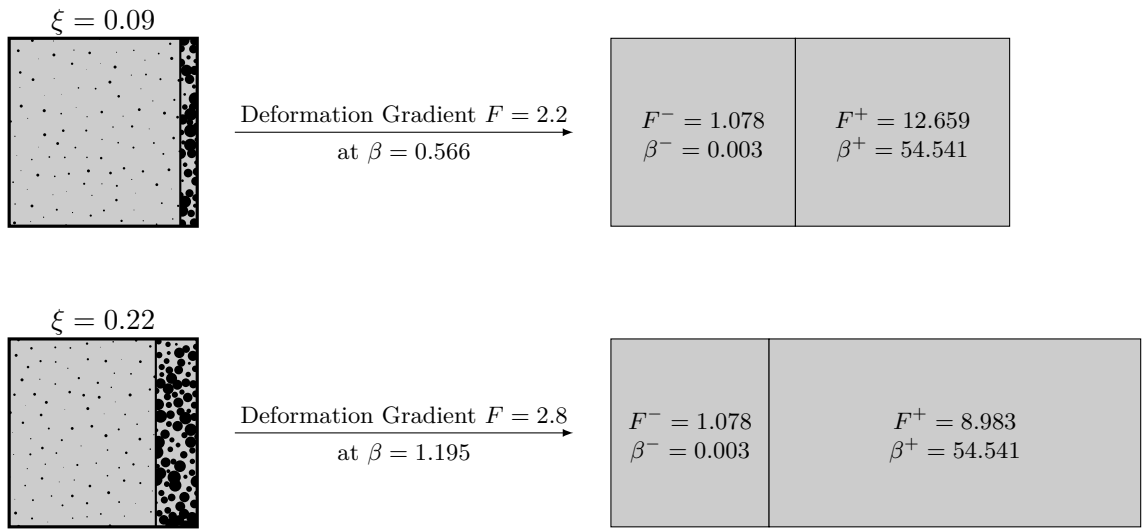
\begin{figure}[t]
    \centering
    \ifthenelse{\boolean{genfig}}
    {
    \tikzsetfigurename{phases}
\begin{tikzpicture}
    \pgfmathsetmacro\squaresize{2.5}
    \pgfmathsetmacro\distance{8.0}
    \pgfmathsetmacro\spacingarrow{0.5}
    \pgfmathsetmacro\spacingvertical{4} 
    \pgfmathsetmacro\volumefractionp{0.09}
    \pgfmathsetmacro\volumefractionm{1-\volumefractionp}
    \pgfmathsetmacro\Fmone{1.078}
    \pgfmathsetmacro\Fpone{12.659}
    \pgfmathsetmacro\lengthminus{\squaresize*\volumefractionm*\Fmone}
    \pgfmathsetmacro\lengthplus{\squaresize*\volumefractionp*\Fpone}
    \pgfmathsetmacro\plusshift{\squaresize*\volumefractionm}
    \pgfmathsetmacro\volumefractionptwo{0.22}
    \pgfmathsetmacro\volumefractionmtwo{1-\volumefractionptwo}
    \pgfmathsetmacro\Fmtwo{1.078}
    \pgfmathsetmacro\Fptwo{8.983}
    \pgfmathsetmacro\lengthminustwo{\squaresize*\volumefractionmtwo*\Fmtwo}
    \pgfmathsetmacro\lengthplustwo{\squaresize*\volumefractionptwo*\Fptwo}
    \pgfmathsetmacro\plusshifttwo{\squaresize*\volumefractionmtwo}

    \edef\minuslist{\poissonpointslist{2.5*\volumefractionm}{2.5}{0.18}{15}}
    \edef\pluslist{\poissonpointslist{2.5*\volumefractionp}{2.5}{0.1}{20}}
    \edef\minuslisttwo{\poissonpointslist{2.5*\volumefractionmtwo}{2.5}{0.18}{15}}
    \edef\pluslisttwo{\poissonpointslist{2.5*\volumefractionptwo}{2.5}{0.1}{20}}
    \coordinate (leftcorner1) at (0,\spacingvertical);
    \coordinate (leftcorner2) at (\distance,\spacingvertical);
    \coordinate (leftcorner3) at (0,0);
    \coordinate (leftcorner4) at (\distance,0);

    \draw[draw=black,fill=gray!40] (leftcorner2) rectangle (\lengthminus+\distance,\squaresize+\spacingvertical) node (minusdescription) [midway,yshift=0.5em] {\smaller $F^-=1.078$};
    \node[below of = minusdescription,yshift=1.5em]{\smaller $\beta^-=0.003$};
    \draw[draw=black,fill=gray!40] (\lengthminus+\distance,\spacingvertical) rectangle (\lengthminus+\lengthplus+\distance,\squaresize+\spacingvertical)node (plusdescription) [midway,yshift=0.5em] {\smaller $F^+=12.659$};
    \node[below of = plusdescription,yshift=1.5em]{\smaller $\beta^+=54.541$};
    \draw[->,-latex] (leftcorner1) ++ (\squaresize+\spacingarrow,\squaresize/2) -- ++ (\distance-\squaresize-\spacingarrow*2,0) node[midway, above] {\smaller Deformation Gradient $F=2.2$} node[midway,below] {\smaller at $\beta = 0.566$};
    \draw[very thick,fill=gray!40] (leftcorner1) rectangle(\squaresize,\squaresize+\spacingvertical) node[above,midway,yshift=3em]{$\xi=\volumefractionp$};

    \draw[draw=black,fill=gray!40] (leftcorner4) rectangle (\lengthminustwo+\distance,\squaresize) node (minusdescription) [midway,yshift=0.5em] {\smaller $F^-=1.078$};
    \node[below of = minusdescription,yshift=1.5em]{\smaller $\beta^-=0.003$};
    \draw[draw=black,fill=gray!40] (\lengthminustwo+\distance,0) rectangle (\lengthminustwo+\lengthplustwo+\distance,\squaresize)node (plusdescription) [midway,yshift=0.5em] {\smaller $F^+=8.983$};
    \node[below of = plusdescription,yshift=1.5em]{\smaller $\beta^+=54.541$};
    \draw[->,-latex] (leftcorner3) ++ (\squaresize+\spacingarrow,\squaresize/2) -- ++ (\distance-\squaresize-\spacingarrow*2,0) node[midway, above] {\smaller Deformation Gradient $F=2.8$} node[midway,below] {\smaller at $\beta = 1.195$};
    \draw[very thick,fill=gray!40] (leftcorner3) rectangle(\squaresize,\squaresize) node[above,midway,yshift=3em]{$\xi=\volumefractionptwo$};

    \draw[thick] (\volumefractionm*\squaresize,\spacingvertical) -- (\volumefractionm*\squaresize,\squaresize+\spacingvertical);
    \draw[thick] (\volumefractionmtwo*\squaresize,0) -- (\volumefractionmtwo*\squaresize,\squaresize);

    \begin{scope}
    \clip (leftcorner1) rectangle (\squaresize,\squaresize+\spacingvertical);
    \foreach \x/\y in \minuslist {
      \pgfmathsetmacro{\radiusminus}{0.1+0.5*rnd}
      \fill (\x,\y+\spacingvertical) circle(\radiusminus pt);
    }
    \clip (\volumefractionm*\squaresize,\spacingvertical) rectangle (\squaresize,\squaresize+\spacingvertical);
    \foreach \x/\y in \pluslist {
      \pgfmathsetmacro{\radiusplus}{0.5+2*rnd}
      \pgfmathsetmacro{\xshifted}{\x+\plusshift}
      \fill (\xshifted,\y+\spacingvertical) circle(\radiusplus pt);
    }
    \end{scope}

    \begin{scope}
    \clip (leftcorner3) rectangle (\squaresize,\squaresize+\spacingvertical);
    \foreach \x/\y in \minuslisttwo {
      \pgfmathsetmacro{\radiusminus}{0.1+0.5*rnd}
      \fill (\x,\y) circle(\radiusminus pt);
    }
    \clip (\volumefractionmtwo*\squaresize,0) rectangle (\squaresize,\squaresize);
    \foreach \x/\y in \pluslisttwo {
      \pgfmathsetmacro{\radiusplus}{0.5+2*rnd}
      \pgfmathsetmacro{\xshifted}{\x+\plusshifttwo}
      \fill (\xshifted,\y) circle(\radiusplus pt);
    }
    \end{scope}
\end{tikzpicture}
    }
    {
    \includegraphics{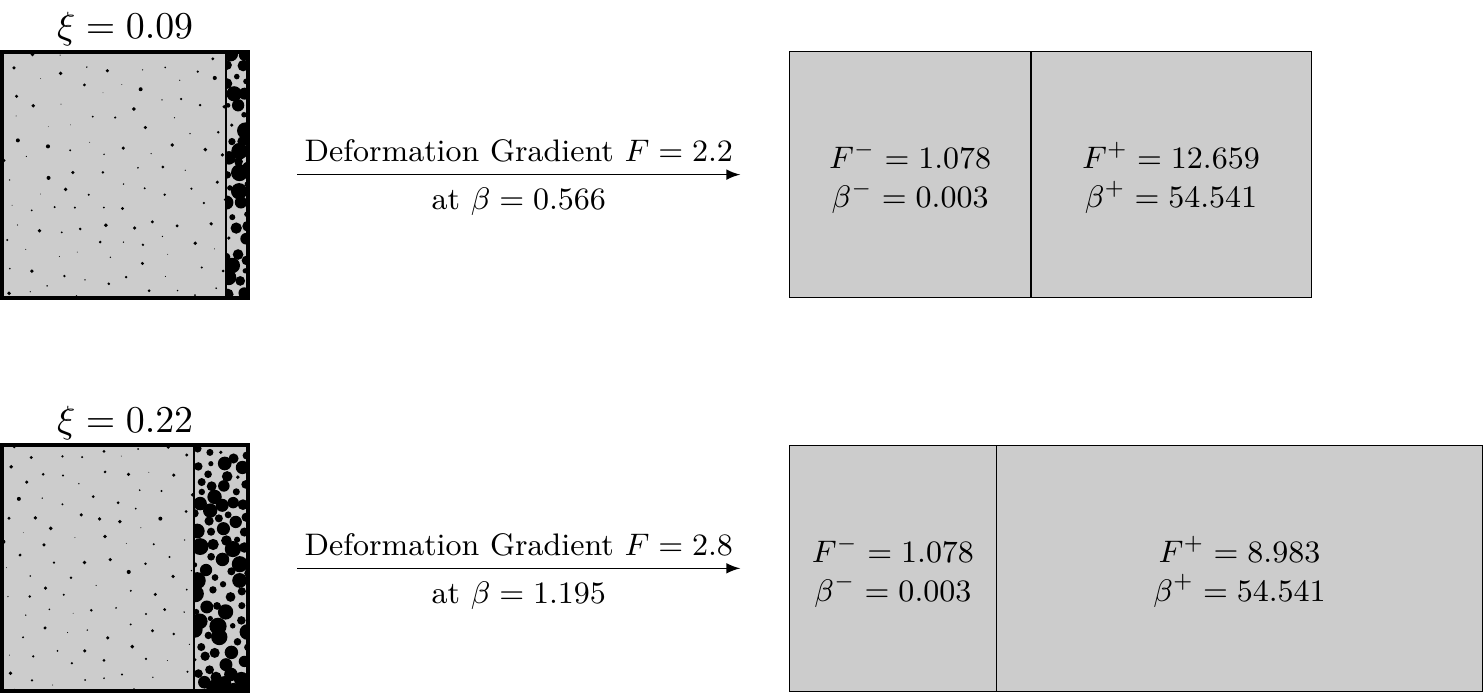}
    }
    \caption{Conceptual sketch of the evolving microstructures.
    Note that the microstructural internal variables $\beta^{\pm}$ and weakly damaged phase deformation gradient $F^-$ stay constant, while $F^+$ can evolve over time.}
    \label{fig:phases}
\end{figure}

In the present contribution, a rather damaged-phase centric perspective is used in order to model the subsequent incremental steps. 
Due to recent contributions in \citet{KohNeuPetPetBal:2021:easa}, the construction of the convex envelope has been significantly accelerated, which opens the possibility to construct a new convex hull in each incremental step. 
Hence, the incremental stress potential~$W(F)$ is \textit{reconvexified}. 
The modeling task consists of setting the internal variable~$\beta$ as well as the properties of $\Xi$ that describe the gradient Young measure, i.e. $F^-, F^+, \beta^-, \beta^+$. 
In Figure~\ref{fig:energy_neohooke}, the incremental stress potential $W(F)$ for a neo-Hookean material is plotted for different states of $\beta$ and their associated incremental steps, respectively. 
Accumulating the behavior over the incremental steps, it can be seen that the non-convex behavior decreases and eventually approaches a convex behavior again. 
This point, where no non-convex behavior is left, marks the state which is mathematically fixated by the material parameter $D_{\infty}$ corresponding to the internal variable state $\beta_\infty$, which is in line with a predefined maximally reachable damage state. 
As soon as the relaxed regime is entered, the weakly damaged phase with deformation gradient $F^- < F$ and strongly damaged phase with $F^+ > F$ exists. 
Ideally, the modeling task is to keep the internal variables $\beta^{\pm}$ fixated, such that $\xi$ allows for an interpretation in terms of damaged state, as well as fulfilling the irreversibility properties of equation \eqref{eq:dissipation-potential}.
This implies that the deformation gradient of the weakly damaged phase $F^-$ is restricted to stay constant, otherwise the internal variable $\beta^-$ would not correspond to $F^-$.
In the novel formulation, the convex hull is constructed in each incremental step.
Then, it is checked whether or not the relaxed regime is entered, if that is the case, the deformation gradients $F^-$ and $F^+$ of the microstructures as well as $\beta^+$ and $\beta^-$ are set.
The internal variable  $\beta$ can be set by the associated $F$ which is obtained by finding the equilibrium of the relaxed incremental stress potential.
Now, in the following subsequent relaxation steps, the convexification is repeated, however, only $F^+$ is allowed to evolve. 
Interestingly, this model changes the point mass $\delta_{F^+}$ over time, but the associated $\beta^+$ remains constant without enforcing it. 
Note that the constant $\beta^+$ is a result of the mathematical structure of the incremental stress potential which marks the asymptotic limit of the damage function.
Thus, only a constant $\beta^-$ and $F^-$ need to be enforced explicitly. 
In Box~\ref{box:materialroutine} an algorithm describing the material routine is given. 
\begin{BOX}[t]
\fbox{\parbox{\textwidth}{
\textbf{Input}: Deformation Gradient $F$, history variables $\beta_k, \beta_k^+, \beta_k^-, F_k^-, F_k^+, \xi_k, d_k, P_k, \mathbb{A}_k$, \\
\ \hspace*{1.25cm} history bool \texttt{first}
\begin{enumerate}[label*=\arabic*.,itemsep=0.05cm]
    \item Initialize return variables $P, \mathbb{A}$
    \item Get $F^-$, $F^+$ from construction of convex envelope $W^C$ according to Algorithm 1 of \cite{KohNeuPetPetBal:2021:easa}
    \item Check if fully damaged state is reached or material still in relaxed regime
    \fbox{\parbox{0.9\textwidth}{\textbf{If} $\beta_k <\beta_k^+$ \textbf{or} $|F^+ - F^-| < \text{tol}$\hfill {\smaller if material not in $D_{\infty}$ damaged state} \\
    \hspace*{1cm} \fbox{\parbox{0.78\textwidth}{ 
            \textbf{If} $W^C(F) \leq W(F)$ \hfill {\smaller relaxed zone}
                \begin{enumerate}[label={},itemsep=0.05cm]
                    \item \textbf{If} \texttt{first} \textbf{then} $F^- = F^-_k$ \hfill {\smaller enforce constant weakly phase}
                    \item compute current volume fraction $\xi = (F - F^-)/(F^+ - F^-)$
                    \item compute current micro bifurcation intensity $d = (F^+ - F^-)/F$
                    \item define $\psi = \psi(F), \psi^+ = \psi(F^+), \psi^- = \psi(F^-), \beta, \beta^+, \beta^-$
                    \item compute first and second derivative $P = \partial_{F} W^C, \mathbb{A} = \partial^2_{FF}W^C$ via automatic differentiation or closed forms in \cite{BalOrt:2012:riv}
                    \item \texttt{first = false}
                \end{enumerate}
            \textbf{Else} \hfill {\smaller convex zone}
            \begin{enumerate}[label={},,itemsep=0.05cm]
                \itemsep0.05em
                \item define $\psi = \psi(F), \beta, \beta^+=\beta, \beta^-=\beta$
                \item compute first and second derivative $P = \partial_{F} W, \mathbb{A} = \partial^2_{FF}W$
                \item \texttt{first = true}
                \item $\xi = \xi_k, d = d_k$
            \end{enumerate}
    }} \\
    \textbf{Else}
            \begin{enumerate}[label={},,itemsep=0.05cm]
            \itemsep0.05em
                \item $P = P_k, \mathbb{A} = \mathbb{A}_k$
            \end{enumerate}
    }}
    \item set history variables $\beta_k = \beta, \beta^+_k = \beta^+, \beta^-_k=\beta^-, F^-_k = F^-, F^+_k = F^+, \xi_k = \xi, d_k = d$
    \item return $P, \mathbb{A}$
\end{enumerate}}}
\caption{Material routine of the novel \textit{reconvexified} model \label{box:materialroutine}}
\end{BOX}
Note that within the material routine Box \ref{box:materialroutine}, the inner boxes do not correspond to loops. 
They are only meant to highlight the important parts of the implementation. 
The volume fraction~$\xi$ evolves now nonlinearly, which is in contrast to \cite{GurMie:2011:edm,BalOrt:2012:riv}. 
In both mentioned publications it has been reported that the volume fraction evolves linearly w.r.t. a linearly growing deformation gradient $F$.\\
A conceptual sketch of the model is given in Figure~\ref{fig:phases}, where a specific scenario is depicted as example. 
The left-hand side of the figure visualizes the mixture of the material according to the volume fraction $\xi$.
Both phases can be associated with a corresponding $\beta$ and thus, a corresponding $D$. 
The right-hand side of the picture shows the associated values which, together with $\xi$, form the entities needed for $\Xi$.
All displayed values in this figure correspond to an actual one-dimensional simulation of a neo-Hookean material.
Note that even though the deformation gradient $F^+$ decreases, the strongly damaged phase has a larger phase fraction
at $F=2.8$ than at $F=2.2$. 
Physically, this model describes distinct features that enhance the model of \cite{GurMie:2011:edm, BalOrt:2012:riv}.
On the one hand, the volume fraction $\xi$ increases non-linearly and thus, more damaged phase is inserted into the mixture. 
On the other hand, due to decreasing $F^+$ and constant $\beta^+$, the strongly damaged phase unloads elastically. 
Since the microstructure mixture increases the fraction of strongly damaged phase constituents, more damaged nuclei are described that need to deform less compared to fewer damaged nuclei. 
These effects lead mathematically to a decrease of the convex hull slope and hence, enables strain softening based on variational modeling. 
Note that irreversibility, i.e. monotonicity of equation \eqref{eq:dissipation-potential} is guaranteed. 
Only one value of the GYM changes in this model and thus, it is likely that a certain minimal transition distance is described. 
However, this is an open mathematical research question. 
The incremental stress potential evolution, as shown in Figure \ref{fig:energy_neohooke} showcases that eventually the function becomes convex again.
In fact, after reaching the $D_{\infty}$ and $\beta_\infty$ state, respectively, only the effective energies $\psi^0$ are active in equation \eqref{eq:isp}.
Thus, the $(1-D)$ model looses its physical meaning, 
since the fully damaged state is described after which a growth condition of $\psi^0$ may increase the function value and its derivatives rapidly. 
Therefore, the new proposed model checks for this state and sets the stresses and tangents constant, such that remaining stresses and tangents are remaining
stresses left in the fully damaged state. 
Note that this is a feature accounting for the often considered numerical demand for robustness that~$D$ should never really become equal to one.
In the following sections, application of this model to one-dimensional, two-dimensional, and three-dimensional simulations is shown, including mesh dependence tests and curve fitting to experimental data. 

\section{Applications and Numerical Examples}
\label{sec:app}
In this section, the novel model is tested in one-dimensional, two-dimensional plane strain, and three-dimensional simulations, where the two- and three-dimensional response is obtained by a microsphere approach. 
First, the general behavior of the one-dimensional model is examined. 
Then the model is adjusted to experimental data to show its applicability to real materials, and finally the model is tested in terms of mesh dependence in four distinct setups. 
All implementations are realized by means of the finite element toolbox Ferrite.jl \cite{CarEkrCon:2021:fj} based on 
the Julia programming language \cite{BezKarShaEde:2012:jfd}.

\subsection{One-dimensional Numerical Analysis
of Reconvexified Model}
\begin{figure}[!t]
    \centering
    \ifthenelse{\boolean{genfig}}
    {
    \tikzsetfigurename{neohooke-stress-deformation}
\begin{tikzpicture}
    \begin{axis}[name=sigmaF,width={0.65\textwidth}, height={5.0cm},
    ylabel={$\sigma$},
    xtick={1,2,3,4,5},
    yticklabel style={
                /pgf/number format/fixed,
                        /pgf/number format/precision=5
    },
    scaled y ticks=false,
    xlabel style={font={\normalsize}}, ylabel style={font={\normalsize}}, legend style={font={\small}}, yticklabel style={font={\normalsize}}, xticklabel style={font={\normalsize}}, axis background/.style={fill={white!89.803921568!black}}, x grid style={white}, y grid style={white}, xmajorgrids, ymajorgrids, legend pos={north east}]
        \addplot+[no markers, color={red!60!gray}, thick]
            table{figures/tikz/rawdata/neohooke_dinfty=0.99_d0=0.5_lambda=0_mu=0.5/sigma_F.txt}
            ;
    \end{axis}
    \begin{axis}[name=xiF,at=(sigmaF.below south west),anchor=above north west,width={0.325\textwidth}, height={4.5cm}, 
    ylabel={$\xi$},
    xtick={1,2,3,4,5},
    xlabel style={font={\normalsize}}, ylabel style={font={\normalsize}}, legend style={font={\small}}, yticklabel style={font={\normalsize}}, xticklabel style={font={\normalsize}}, axis background/.style={fill={white!89.803921568!black}}, x grid style={white}, y grid style={white}, xmajorgrids, ymajorgrids, legend pos={north east}]
        \addplot+[no markers, color={red!60!gray}, thick]
            table{figures/tikz/rawdata/neohooke_dinfty=0.99_d0=0.5_lambda=0_mu=0.5/xi_F.txt}
            ;
    \end{axis}
    \hspace{0.2cm}
    \begin{axis}[name=dF,at=(xiF.right of south east),anchor=left of south west,width={0.325\textwidth}, height={4.5cm}, 
    ylabel={$d$},
    xtick={1,2,3,4,5},
    xlabel style={font={\normalsize}}, ylabel style={font={\normalsize},yshift=-0.2cm}, legend style={font={\small}}, yticklabel style={font={\normalsize}}, xticklabel style={font={\normalsize}}, axis background/.style={fill={white!89.803921568!black}}, x grid style={white}, y grid style={white}, xmajorgrids, ymajorgrids, legend pos={north east}]
        \addplot+[no markers, color={red!60!gray}, thick]
            table{figures/tikz/rawdata/neohooke_dinfty=0.99_d0=0.5_lambda=0_mu=0.5/d_F.txt}
            ;
    \end{axis}
    \node at (3.8,-4.3) {\large Deformation Gradient $F$};
\end{tikzpicture}
    }
    {
    \includegraphics{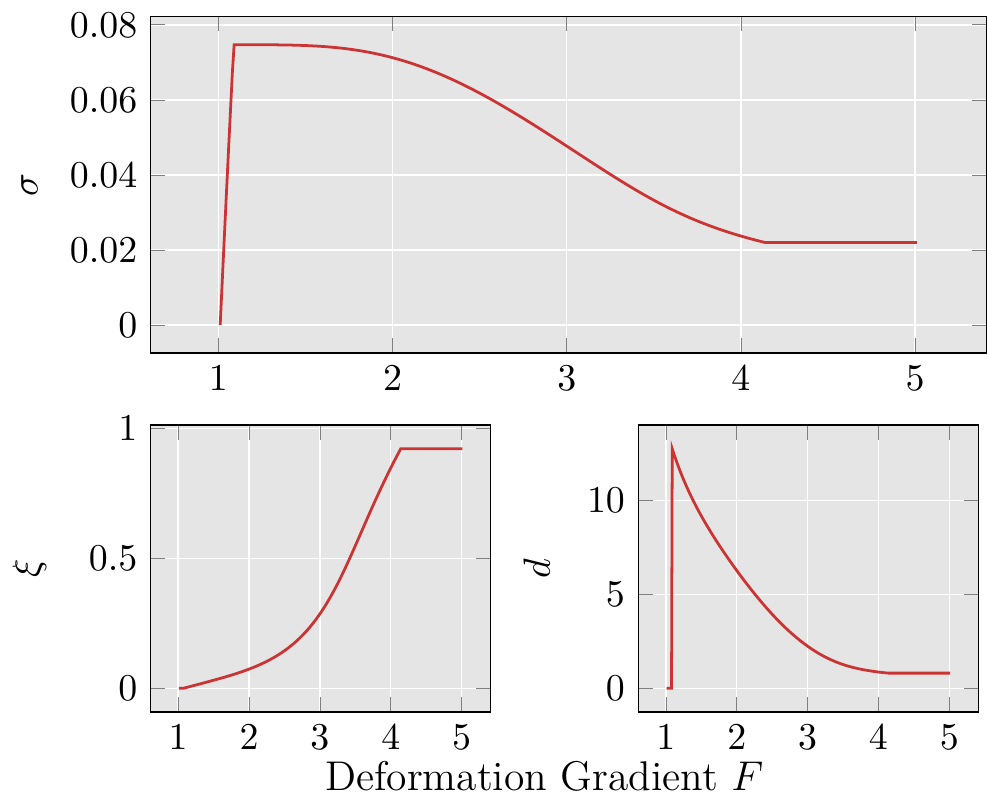}
    }
    \caption{Cauchy stress vs. deformation gradient of reconvexified model using neo-Hooke energy density (top), evolution of volume fraction~$\xi$ (bottom left), and microbifurcation intensity~$d$ (bottom right).
    Significant strain-softening as well as a non-linear behavior of~$\xi$ can be observed.}
    \label{fig:neohooke}
\end{figure}
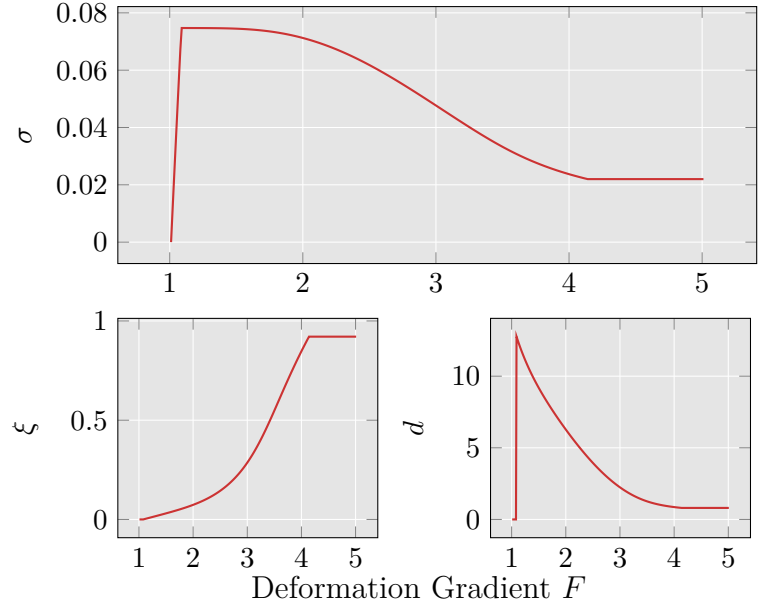

\begin{figure}[!t]
    \centering
    \ifthenelse{\boolean{genfig}}
    {
    \tikzsetfigurename{yeoh-stress-deformation}
\begin{tikzpicture}
    \begin{axis}[name=sigmaF,width={0.65\textwidth}, height={5cm},
    ylabel={$\sigma$},
    xtick={1.0,1.2,1.4,1.6,1.8},
    yticklabel style={
                /pgf/number format/fixed,
                        /pgf/number format/precision=5
    },
    scaled y ticks=false,
    xlabel style={font={\normalsize}}, ylabel style={font={\normalsize}}, legend style={font={\small}}, yticklabel style={font={\normalsize}}, xticklabel style={font={\normalsize}}, axis background/.style={fill={white!89.803921568!black}}, x grid style={white}, y grid style={white}, xmajorgrids, ymajorgrids, legend pos={north east}]
        \addplot+[no markers, color={red!60!gray}, thick]
            table{figures/tikz/rawdata/yeoh_dinfty=0.99_d0=1.0_c1=6.0_c2=1.0_c3=900.0/sigma_F.txt}
            ;
    \end{axis}
    \begin{axis}[name=xiF,at=(sigmaF.below south west),anchor=above north west,width={0.325\textwidth}, height={4.5cm}, 
    ylabel={$\xi$},
    xtick={1.0,1.2,1.4,1.6,1.8},
    xlabel style={font={\normalsize}}, ylabel style={font={\normalsize}}, legend style={font={\small}}, yticklabel style={font={\normalsize}}, xticklabel style={font={\normalsize}}, axis background/.style={fill={white!89.803921568!black}}, x grid style={white}, y grid style={white}, xmajorgrids, ymajorgrids, legend pos={north east}]
        \addplot+[no markers, color={red!60!gray}, thick]
            table{figures/tikz/rawdata/yeoh_dinfty=0.99_d0=1.0_c1=6.0_c2=1.0_c3=900.0/xi_F.txt}
            ;
    \end{axis}
    \hspace{0.2cm}
    \begin{axis}[name=dF,at=(xiF.right of south east),anchor=left of south west,width={0.325\textwidth}, height={4.5cm}, 
    ylabel={$d$},
    xtick={1.0,1.2,1.4,1.6,1.8},
    ytick={0,0.15,0.3},
    xlabel style={font={\normalsize}}, ylabel style={font={\normalsize},yshift=-0.2cm}, legend style={font={\small}}, yticklabel style={font={\normalsize}}, xticklabel style={font={\normalsize}}, axis background/.style={fill={white!89.803921568!black}}, x grid style={white}, y grid style={white}, xmajorgrids, ymajorgrids, legend pos={north east}]
        \addplot+[no markers, color={red!60!gray}, thick]
            table{figures/tikz/rawdata/yeoh_dinfty=0.99_d0=1.0_c1=6.0_c2=1.0_c3=900.0/d_F.txt}
            ;
    \end{axis}
    \node at (3.8,-4.3) {\large Deformation Gradient $F$};
\end{tikzpicture}
    }
    {
    \includegraphics{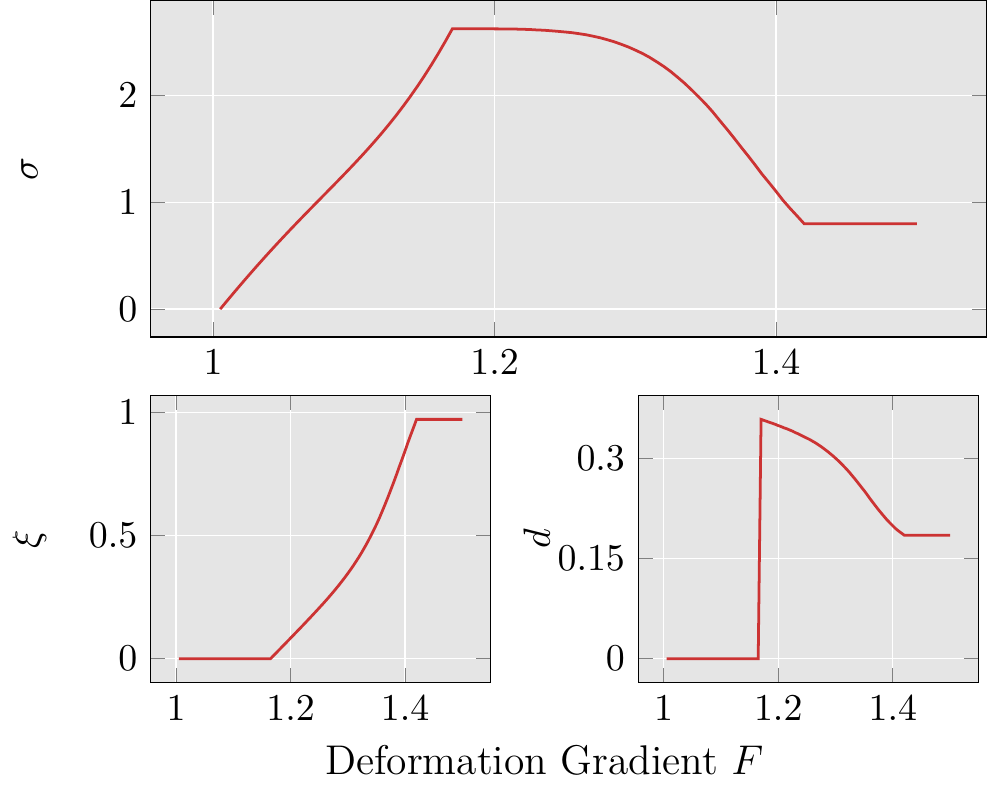}
    }
    \caption{
    Cauchy stress vs. deformation gradient of reconvexified model using Yeoh energy density (top), evolution of volume fraction~$\xi$ (bottom left) and microbifurcation intensity~$d$ (bottom right). 
    Note the larger regime of regular damage evolution until approximately $F=1.18$ and subsequent strain-softening due to the reconvexified model.}
    \label{fig:yeoh}
\end{figure}
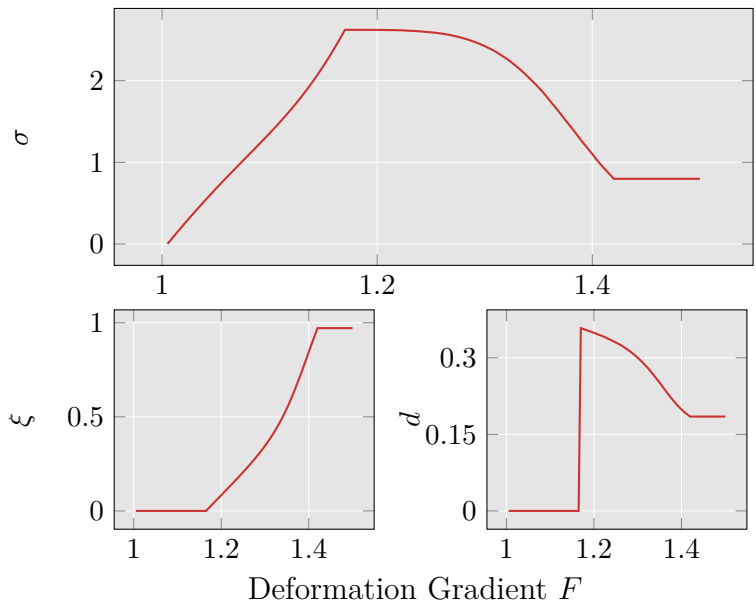

In the previous section, it was mentioned how the variational modeling leads to two unique features of the new reconvexified formulation. 
The non-linear evolution of~$\xi$ as well as the decreasing slope of the convex hull was stated which will be demonstrated 
in the following. 
To this end, the neo-Hookean effective energy~$\psi^0$ 
%
\eb
\psi_{\text{NH}}^0(\bC) \defeq \frac{\mu}{2} (I_1 - 3) - \mu \ln(J) + \frac{\lambda}{2} \ln(J)^2,
\ee
with $J=\sqrt{I_3}$, is considered. 
For the one-dimensional case 
we define
$\bF = \text{diag}[F, 1, 1]$, where~$F$ denotes the contribution of the deformation gradient in the axis of the one-dimensional problem. 
The material response is visualized in Figure~\ref{fig:neohooke} for the material parameters $\lambda = 0.0, \mu = 0.5, D_0 = 0.5$, and $D_\infty=0.99$. 
Note that~$\lambda$, $\mu$, and $D_0$ share the same units of energy density, $D_\infty$ is unitless. 
Specifically, the Cauchy stress in the same unit as the material parameters is plotted versus the deformation gradient for all incremental steps in Figure~\ref{fig:energy_neohooke} (top). 
%
As can be seen, the stress increases in the first incremental steps while undergoing regular damage evolution until the relaxed regime is entered. 
There, the stress shows the desired strain softening, which can be observed by 
the incrementally decreasing slope of the subsequent~$W^C(F)$ after convexifying $W(F)$. 
Remember that~$W(F)$ changes due to the 
repeated minimization of $\mathcal{W}(F,\beta)$ w.r.t. $\beta$ and associated convexification in every incremental step. 
In this part of the figure, at around $F=4.1$, the fully damaged state is reached and thus, the stress remains constant. 
At the bottom of the figure, the evolution of the volume fraction~$\xi$ as well as the microbifurcation intensity~$d$ is plotted. 
The non-linear evolution of~$\xi$ can be seen, which is in contrast to the linear response from \cite{GurMie:2011:edm,BalOrt:2012:riv}. 
In this plot, the irreversibility of the evolution~$\xi$ can be observed as well. 
On the other hand, the evolution of microbifurcation intensity~$d$ shows a similar behavior as in \cite{BalOrt:2012:riv}.\\ 
Since the regime which does not require relaxation
is rather small in the neo-Hookean case, a Yeoh-type material is tested as well. 
The effective strain energy density~$\psi^0$ is defined as 
\eb
\label{eq:yeoh}
\psi_{\text{Yeoh}}^0(\bC) \defeq c_1 \left(I_1 I_3^{-1/3}-3\right) + c_2 \left(I_1 I_3^{-1/3}-3\right)^2 + c_3 \left(I_1 I_3^{-1/3}-3\right)^3.
\ee
Here, the material parameters were chosen to $c_1=6.0,c_2=1.0,c_3=900,D_0=1.0$, and $D_\infty=0.99$. 
The parameters $c_1$, $c_2$, $c_3$, and $D_0$ share the same energy density units as the stress.
In Figure~\ref{fig:yeoh}, the material response in terms of a $\sigma$--F plot as well as the evolution of the microstructural variables~$\xi$ and~$d$ is shown. 
Again, the distinct features of the novel model can be seen in the microstructural variables, but more importantly, the stresses evolve as observed in soft materials. 
A relatively large unrelaxed regime until approximately $F=1.18$ can be seen, where regular damage evolution takes place, after which the reconvexified model is activated. 
In the relaxed regime, a gradual reduction of the stress with increasing strain, i.e. strain softening, can be observed. 
The practically more relevant behavior obtained for the Yeoh energy motivates to fit the model to real experimental data of soft materials as e.g., soft biological tissues to check if also real materials can be well represented.

\subsubsection{Adjustment to Experiments}
\begin{figure}[!t]
    \begin{minipage}[t]{0.35\textwidth}
            \vspace{7.3em}
        \begin{table}[H]
            \centering
            \begin{tabular}{lllll}
                \hline
                $D_0$ & $D_\infty$ & $c_1$ & $c_2$ & $c_3$ \\ \hline
                12.5  & 0.88       & 50    & 200   & 20000 \\
                \hline
            \end{tabular}
            \vspace{3.5em}
            \caption{Material parameters obtained from handfitting.
            \label{tab:handfit}}
        \end{table}
    \end{minipage}
    \begin{minipage}[t]{0.63\textwidth}
        \begin{figure}[H]
            \centering
            \ifthenelse{\boolean{genfig}}
            {
            \tikzsetfigurename{experiment-fit}
\begin{tikzpicture}
    \begin{axis}[name=sigmaF,width={0.99\textwidth}, height={5.5cm},
    ylabel={$\sigma$},
    xlabel={$\varepsilon=\frac{\Delta l}{l_0}$},
    xtick={0,0.1,0.2,0.3},
    yticklabel style={
                /pgf/number format/fixed,
                        /pgf/number format/precision=5
    },
    scaled y ticks=false,
    xlabel style={font={\normalsize}}, ylabel style={font={\normalsize}}, legend style={font={\small}}, yticklabel style={font={\normalsize}}, xticklabel style={font={\normalsize}}, axis background/.style={fill={white!89.803921568!black}}, x grid style={white}, y grid style={white}, xmajorgrids, ymajorgrids, legend pos={south east}]
        \addplot+[color={black}, thick]
            table[col sep=comma]{figures/tikz/rawdata/aaa_long.csv}
            ;
        \addlegendentry{AAA$_{\text{long}}$}
        \addplot+[no markers, color={red!60!gray}, ultra thick]
            table{figures/tikz/rawdata/AAA_long_handfit_yeoh.txt}
            ;
        \addlegendentry{Model}
    \end{axis}
\end{tikzpicture}
            }
            {
            \includegraphics{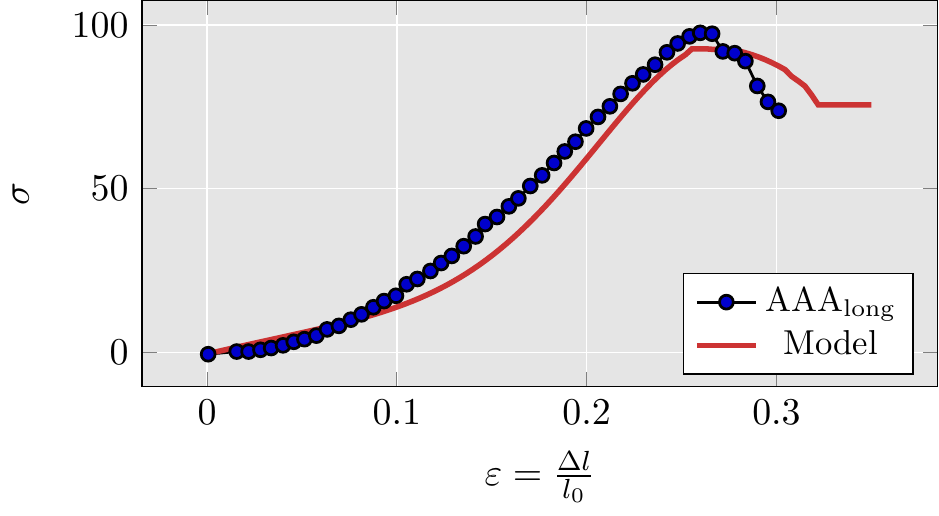}
            }
            \caption{
            Comparison in terms of stress vs. strain of reconvexified model response using handfitted parameters with experimental data for tissue sample from abdominal aortic aneurysm from~\cite{RagWebVor:1996:evb}. 
            As can be seen, the agreement is already reasonable.}
            \label{fig:handfit}
        \end{figure}
    \end{minipage}
\end{figure}

The practical relevance of the proposed model is tested by fitting the parameters to real experimental data. 
The Yeoh effective energy $\psi^0_{\text{Yeoh}}$ is used. 
Thus, the model contains 5 independent material parameters, which need to be adjusted, i.e. $D_0, D_\infty$ (cf. equation \eqref{eq:damagefunction}), $c_1, c_2,$ and $c_3$ (cf. equation \eqref{eq:yeoh}). 
Data from~\citet{RagWebVor:1996:evb} for the mechanical response of tissue samples taken from an abdominal aortic aneurysm (AAA) is transferred by reading from the figures of the aforementioned publication, since no formatted text based data is publicly available. 
Then the material parameters were changed by hand until a reasonable agreement of the model response and the experiment was achieved. 
Here, hand fitting was chosen to show that a realistic response can be obtained even without applying a sophisticated parameter optimization scheme. 
However, if such an automated procedure is desired, additional constraints addressing an appropriate convexification grid for changing material parameter values need to be incorporated in the optimization scheme. 
This is due to the fact that a change in the parameters also results in a change of the relaxed regime.
However, already a fit by hand shows a reasonable agreement of the model response with the experimental data, as can be seen in Figure~\ref{fig:handfit}. 
The obtained parameters are given in Table~\ref{tab:handfit}. 
Admittedly, the material parameter values are debatable as well as the rather high remaining stress of the $D_\infty$ state beginning at approximately $\varepsilon=0.32$. 
However, the results show that in principle
the proposed approach 
is capable of showing the strain softening
behavior of soft biological tissues. 

\subsubsection{Stress Softening}
\begin{figure}[h]
    \centering
    \ifthenelse{\boolean{genfig}}
    {
    \tikzsetfigurename{stress-softening}
\begin{tikzpicture}
    \begin{axis}[name=sigmaF,width={0.65\textwidth}, height={5.0cm},
    ylabel={$\sigma$},
    xtick={1.0,1.25,1.5,1.75},
    yticklabel style={
                /pgf/number format/fixed,
                        /pgf/number format/precision=5
    },
    scaled y ticks=false,
    xlabel style={font={\normalsize}}, ylabel style={font={\normalsize}}, legend style={font={\small}}, yticklabel style={font={\normalsize}}, xticklabel style={font={\normalsize}}, axis background/.style={fill={white!89.803921568!black}}, x grid style={white}, y grid style={white}, xmajorgrids, ymajorgrids, legend pos={north east}]
        \addplot+[no markers, color={red!60!gray}, thick]
            table{figures/tikz/rawdata/stress-softening-yeoh_1.0_0.2_2.0_pull=0.75_5_cycles_sigma_F.txt}
            ;
    \end{axis}
    \begin{axis}[name=xiF,at=(sigmaF.below south west),anchor=above north west,width={0.325\textwidth}, height={4.5cm}, 
    ylabel={$\xi$},
    xtick={1.0, 1.25,1.5,1.75},
    xlabel style={font={\normalsize}}, ylabel style={font={\normalsize}}, legend style={font={\small}}, yticklabel style={font={\normalsize}}, xticklabel style={font={\normalsize}}, axis background/.style={fill={white!89.803921568!black}}, x grid style={white}, y grid style={white}, xmajorgrids, ymajorgrids, legend pos={north east}]
        \addplot+[no markers, color={red!60!gray}, thick]
            table{figures/tikz/rawdata/stress-softening-yeoh_1.0_0.2_2.0_pull=0.75_5_cycles_xi_F.txt}
            ;
    \end{axis}
    \begin{axis}[name=dF,at=(xiF.right of south east),anchor=left of south west,width={0.325\textwidth}, height={4.5cm}, 
    ylabel={$d$},
    xtick={1.0,1.25,1.5,1.75},
    xlabel style={font={\normalsize}}, ylabel style={font={\normalsize},yshift=-0.2cm}, legend style={font={\small}}, yticklabel style={font={\normalsize}}, xticklabel style={font={\normalsize}}, axis background/.style={fill={white!89.803921568!black}}, x grid style={white}, y grid style={white}, xmajorgrids, ymajorgrids, legend pos={north east}]
        \addplot+[no markers, color={red!60!gray}, thick]
            table{figures/tikz/rawdata/stress-softening-yeoh_1.0_0.2_2.0_pull=0.75_5_cycles_d_F.txt}
            ;
    \end{axis}
    \node at (3.8,-4.3) {\large Deformation Gradient $F$};
\end{tikzpicture}
    }
    {
    \includegraphics{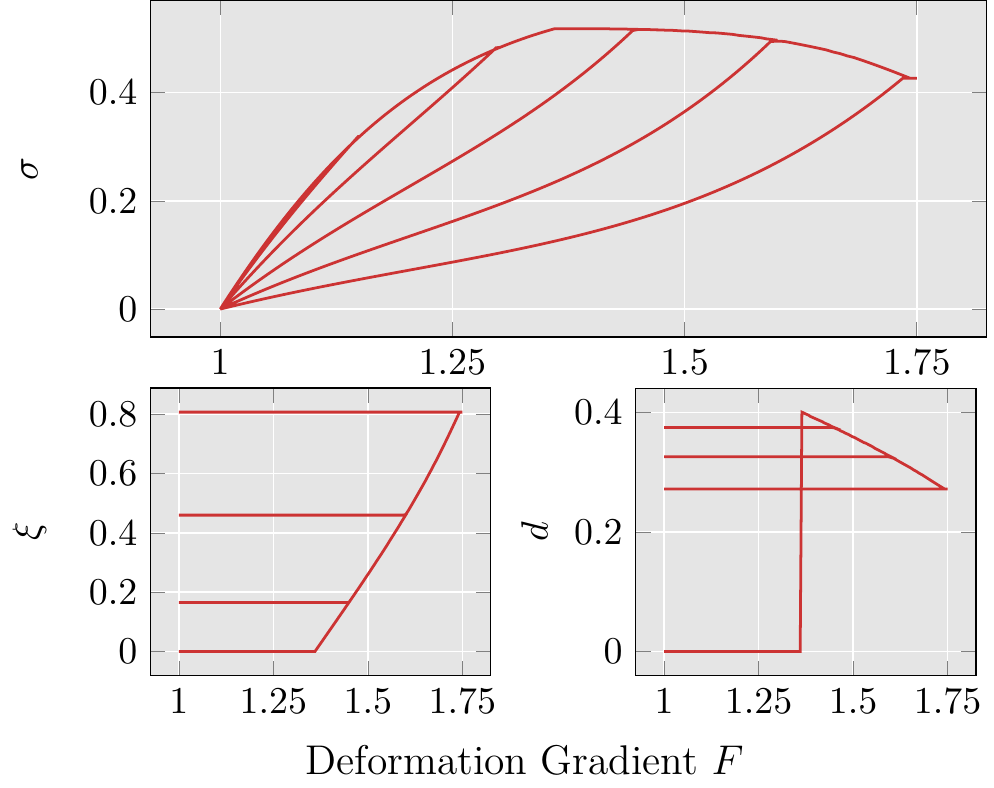}
    }
    \caption{Model response for cyclic loading using a Yeoh base material: evolution of stress (top), volume fraction of strongly damaged phase (bottom left), and microbifurcation intensity (bottom right) vs. deformation gradient. 
    Clearly, stress- as well as strain softening can be described.}
    \label{fig:stress-softening}
\end{figure}

The modeling of stress softening for the novel reconvexified material description is achieved by following the approach of \cite[Section 4.1]{SchBal:2016:riv}. 
In case of unloading within the convexified regime the convexified incremental stress potential $W^C(F)$ is substituted by a modified polyconvex hyperelastic energy $\tilde{W}(F)$
\eb
W^C \Leftarrow \tilde{W}(F,\eta) \defeq \eta \psi^0(F) \qquad \text{with} \qquad P = \eta \frac{\partial \psi^0(F)}{\partial F} \qquad \text{and} \qquad \mathbb{A} = \eta \frac{\partial^2 \psi^0(F)}{\partial F \partial F}.
\ee
%
The material response of the stress-softening approach in combination with the reconvexified model can be seen in Figure~\ref{fig:stress-softening} for a Yeoh effective energy density with the material parameters $c_1=1.0$, $c_2=0.2$, $c_3=2.0$ and damage function parameters $D_0 = 0.3$, $D_\infty = 0.9$. 
As can be observed, stress- as well as strain softening can be well described in principle. 

\subsubsection{Mesh Independence of One-dimensional Reconvexified Model
\label{sec:meshind-one}}
\begin{figure}[t]
     \centering
     \begin{minipage}{0.1\textwidth}
         \hspace{2cm}
        \ifthenelse{\boolean{genfig}}
        {
         \tikzsetfigurename{perturbation-problem}
\begin{tikzpicture}
    \pgfmathsetmacro{\cubex}{0.6}
    \pgfmathsetmacro{\cubey}{4.5}
    \pgfmathsetmacro{\cubez}{0.6}
    \pgfmathsetmacro{\distance}{0.2}
    \pgfmathsetmacro{\shift}{0.8}
    \pgfmathsetmacro{\linelength}{0.2}
    \pgfmathsetmacro{\inclination}{0.15}
    \pgfmathsetmacro{\shiftz}{0.1}
    \pgfmathsetmacro{\nodesize}{4.0}
    \pgfmathsetmacro{\Dshift}{1.2}
    \pgfmathsetmacro{\arrowlength}{1.0}
    \draw[black,fill=gray!40] (0,0,0) -- ++(-\cubex,0,0) -- ++(0,-\cubey,0) node[midway,above,rotate=90,xshift=-\Dshift cm]{\smaller $D_{\infty}=0.99$} node[midway,above,rotate=90,xshift=\Dshift cm]{\smaller $D_{\infty}=0.99 - \epsilon$} -- ++(\cubex,0,0) -- cycle;
    \draw[black,fill=gray!40] (0,0,0) -- ++(0,0,-\cubez) -- ++(0,-\cubey,0) -- ++(0,0,\cubez) -- cycle;
    \draw[black,fill=gray!40] (0,0,0) -- ++(-\cubex,0,0) -- ++(0,0,-\cubez) -- ++(\cubex,0,0) -- cycle;
    \draw[black] (-\cubex,-\cubey/2,0) -- (0,-\cubey/2,0) -- (0,-\cubey/2,-\cubez);

    \draw[black] (\shift*0.8,-\cubey/2,0) -- (\shift*1.2,-\cubey/2,0);
    \draw[black] (\shift*0.8,-\cubey,0) -- (\shift*1.2,-\cubey,0);
    \draw[black] (\shift*0.8,0,0) -- (\shift*1.2,0,0);

    \draw[black] (\shift*0.8,-\cubey/2,-\shiftz) -- (\shift*1.2,-\cubey/2,\shiftz);
    \draw[black] (\shift*0.8,-\cubey,-\shiftz) -- (\shift*1.2,-\cubey,\shiftz);
    \draw[black] (\shift*0.8,0,-\shiftz) -- (\shift*1.2,0,\shiftz);

    \draw[dashed] (0,-\cubey/2,-\cubez) -- (-\cubex, -\cubey/2, -\cubez) -- (-\cubex, -\cubey/2, 0);
    \draw[dashed] (0,-\cubey,-\cubez) -- (-\cubex, -\cubey, -\cubez) -- (-\cubex, -\cubey, 0);

    \foreach \x in {0,-\distance,...,-\cubex}
        \draw[black] (\x,-\cubey,0) -- (\x,-\cubey-\linelength,\inclination) ;
    \foreach \x in {0,-\distance,...,-\cubex}
        \draw[black] (0,-\cubey,\x) -- (\inclination,-\cubey-\linelength,\x) ;

    \draw[->,-latex] (-\cubex/2,0,-\cubez/2) -- (-\cubex/2,\arrowlength,-\cubez/2) node[right,xshift=0.2em]{$u_D$};
    \draw[] (\shift,0,0) -- (\shift,-\cubey/2,0) node[midway,below,rotate=90]{\smaller $(1-\kappa)L$} -- (\shift,-\cubey,0) node[midway,below,rotate=90]{\smaller $\kappa L$};
    \draw[thick] (-\cubex/2,0,-\cubez/2) -- (-\cubex/2,-\cubey/2,-\cubez/2) node[circle,fill=red!40!gray, inner sep=0pt, minimum size=\nodesize pt,pos=0.0]{ } -- (-\cubex/2,-\cubey,-\cubez/2) node[circle,fill=blue!40!gray, inner sep=0pt, minimum size=\nodesize pt,pos=0.0]{ } node[circle,fill=black, inner sep=0pt, minimum size=\nodesize pt,pos=1.0]{ };
\end{tikzpicture}
        }
        {
        \includegraphics{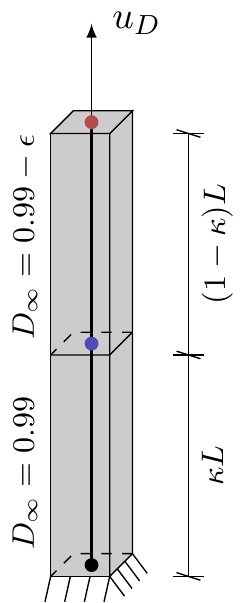}
        }
     \end{minipage}
     \begin{minipage}{0.76\textwidth}
         \centering
        \ifthenelse{\boolean{genfig}}
        {
         \input{figures/tikz/perturbation-test}
        }
        {
         \includegraphics{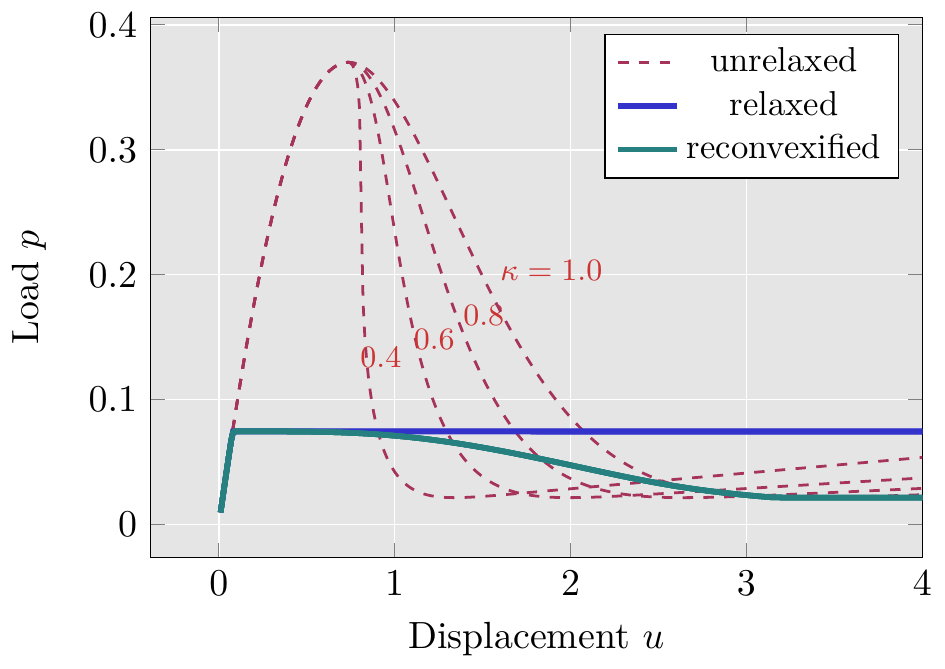}
        }
     \end{minipage}
     \caption{The two element material perturbation problem is sketched on the left-hand side with the associated force-displacement diagram on the right-hand side for the unrelaxed, relaxed \cite{BalOrt:2012:riv} and proposed reconvexified model. 
     Clearly, the relaxed formulations lead to mesh-independent results. 
     However, in contrast to the model in~\cite{BalOrt:2012:riv}, the proposed reconvexification approach enables the description of strain-softening.}
     \label{fig:perturbation}
\end{figure}

The one-dimensional model is tested in terms of mesh independence by the two element material perturbation test that was introduced in the relaxed plasticity setting in \citet{LamMieDet:2003:ern}. 
Within the setting for relaxed variational damage formulation, 
\citet{GurMie:2011:edm} and \citet{BalOrt:2012:riv} used the test in the small and finite strain case, respectively. 
Here, the neo-Hooke effective energy~$\psi^0_{\text{NH}}$ with material parameters $\lambda = 0.0$, $\mu = 0.5$, $D_0=0.5$, and $D_\infty=0.99$ is used, such that a comparison to \cite{BalOrt:2012:riv} is possible. 
The test consists of two one-dimensional elements with linear basis functions. 
So, in total, the domain is discretized with three nodes, two prescribed by Dirichlet boundary conditions. 
One node, the lowest in Figure~\ref{fig:perturbation}, is fixed, i.e. a homogeneous Dirichlet boundary condition is applied. 
The middle node in this figure has a true degree of freedom, while the most upper node is prescribed by a heterogeneous, linearly increasing incremental Dirichlet boundary condition. 
In this problem set up, the domain size is fixed to $L=1$, where the individual element sizes are parameterized by $\kappa L$ and $(1-\kappa)L$ for the lower and upper element, respectively. 
Now, the upper element material parameter $D_\infty$ is distorted by a small perturbation $\epsilon$. 
The perturbation is small enough such that it has no physical meaning, however, it is sufficient to trigger the well-known mesh dependence of the standard (unrelaxed)
damage 
formulation. 
In the simulation, $\epsilon$ was set to $10^{-8}$. 
On the right-hand side of Figure~\ref{fig:perturbation}, the force-displacement curve can be seen, where the unrelaxed formulation is plotted in purple, the relaxed formulation of~\cite{BalOrt:2012:riv} in blue, and the novel reconvexified formulation in teal.
The characteristic mesh dependence of the unrelaxed formulation can be observed
while the relaxed and reconvexified formulation lead to mesh-independent results, i.e. all curves for $\kappa \in \{0.4,0.6,0.8,1.0\}$ coincide. 
Also, the problem set up was tested in a force- instead of displacement-driven setting with an Arc length solver (\citet{Cri:1981:fii}). 
There, the same behavior was observed. 
\subsection{Analysis of Microsphere Approach based on Reconvexified Model}
The generalization of one-dimensional material models to three-dimensional responses can be achieved by the procedure reported in \citet{FreIhl:2010:gom}. 
The aforementioned contribution presents an approach that is mathematically similar to the microsphere approach, which dates back to \citet{BazOh:1986:eni,MieGokLul:2004:mar}. 
In the microsphere model, the one-dimensional material law is evaluated in microfibers (or representative directions) and multiplied with suitable structural tensors. 
The elongation of the $\alpha$-th fiber is denoted by $F^{\alpha} = |\bF \bA_{\alpha}|$, where $\bA \defeq \bA(\varphi,\vartheta)$ denotes the direction of the fiber, which depends on the spherical angles~$\varphi$ and~$\vartheta$. 
The one-dimensional material law is evaluated in all ${\alpha}$ directions
\eb
W^{\alpha}\defeq W(F^{\alpha}), \qquad P^{\alpha} \defeq \pp{W^{\alpha}}{F^{\alpha}}.
\ee
Now, all important one-dimensional model quantities, such as stresses and tangent moduli, are integrated over the unit sphere, in order to obtain the three-dimensional first Piola-Kirchhoff stress and the nominal tangent moduli, (cf. \cite{FreIhl:2010:gom}), by
\eb
\bP = \int_0^{2\pi} \int_0^{\pi} \rho P^{\alpha} \bA \otimes \bA \sin \vartheta \text{ d}\vartheta \text{ d} \varphi \qquad
\mathbb{A} = \int_0^{2\pi} \int_0^{\pi} \rho \pp{P^{\alpha}}{F^{\alpha}} \bA \otimes \bA \otimes \bA \otimes \bA \sin \vartheta \text{ d}\vartheta \text{ d} \varphi .
\ee
Here, $\rho$ denotes an orientation distribution function, which we restrict to a uniform distribution in this publication. 
Note that more sophisticated distribution functions can be used to realistically describe materials with dispersed fibers such as soft biological tissues, cf. \citet{SchBal:2016:riv}. 
Since all fiber evaluations are independent of each other, a trivial parallelization is utilized, such that each fiber is evaluated on a designated thread on the CPU. 
The parallelization is achieved with the FLoops.jl \cite{Ara:2022:fjf} library. 
The question arises if the integration of one-dimensional convex functions yields again a convex energy density in 3D. 
In section~\ref{sec:meshind-one} it was shown that one single fiber leads to the expected mesh-independent results. 
Previous numerical analysis has shown that non-convexities, which may arise from the microsphere integration between the individual fiber directions, tend to only show negligible effects on the overall mesh independence, cf.~\citet{SchBal:2016:riv}, \citet{KohNeuPetPetBal:2021:easa}. 
However, for the reconvexified approach proposed here, it is not a priori clear that the influence of small non-convexities between the fiber directions on the overall mesh independence of numerical simulations is indeed small. 
Thus, in the following, the mesh independence of the microsphere-integrated reconvexified model is examined.

\subsubsection{Two Element Material Perturbation}
\begin{figure}[t]
     \centering
     \hspace{1cm}
     \begin{minipage}{0.1\textwidth}
         \hspace{2cm}
        \ifthenelse{\boolean{genfig}}
        {
         \tikzsetfigurename{perturbation-problem3D}
\begin{tikzpicture}
    \pgfmathsetmacro{\cubex}{0.6}
    \pgfmathsetmacro{\cubey}{4.5}
    \pgfmathsetmacro{\cubez}{0.6}
    \pgfmathsetmacro{\distance}{0.2}
    \pgfmathsetmacro{\shift}{0.8}
    \pgfmathsetmacro{\linelength}{0.2}
    \pgfmathsetmacro{\inclination}{0.15}
    \pgfmathsetmacro{\shiftz}{0.1}
    \pgfmathsetmacro{\nodesize}{4.0}
    \pgfmathsetmacro{\Dshift}{1.2}
    \pgfmathsetmacro{\arrowlength}{1.0}
    \draw[black,fill=gray!40] (0,0,0) -- ++(-\cubex,0,0) -- ++(0,-\cubey,0) node[midway,above,rotate=90,xshift=-\Dshift cm]{\smaller $D_{\infty}=0.99$} node[midway,above,rotate=90,xshift=\Dshift cm]{\smaller $D_{\infty}=0.99-\epsilon$} -- ++(\cubex,0,0) -- cycle;
    \draw[black,fill=gray!40] (0,0,0) -- ++(0,0,-\cubez) -- ++(0,-\cubey,0) -- ++(0,0,\cubez) -- cycle;
    \draw[black,fill=gray!40] (0,0,0) -- ++(-\cubex,0,0) -- ++(0,0,-\cubez) -- ++(\cubex,0,0) -- cycle;
    \draw[black] (-\cubex,-\cubey/2,0) -- (0,-\cubey/2,0) -- (0,-\cubey/2,-\cubez);

    \draw[black] (\shift*0.8,-\cubey/2,0) -- (\shift*1.2,-\cubey/2,0);
    \draw[black] (\shift*0.8,-\cubey,0) -- (\shift*1.2,-\cubey,0);
    \draw[black] (\shift*0.8,0,0) -- (\shift*1.2,0,0);

    \draw[black] (\shift*0.8,-\cubey/2,-\shiftz) -- (\shift*1.2,-\cubey/2,\shiftz);
    \draw[black] (\shift*0.8,-\cubey,-\shiftz) -- (\shift*1.2,-\cubey,\shiftz);
    \draw[black] (\shift*0.8,0,-\shiftz) -- (\shift*1.2,0,\shiftz);

    \draw[dashed] (0,-\cubey/2,-\cubez) -- (-\cubex, -\cubey/2, -\cubez) -- (-\cubex, -\cubey/2, 0);
    \draw[dashed] (0,-\cubey,-\cubez) -- (-\cubex, -\cubey, -\cubez) -- (-\cubex, -\cubey, 0);

    \foreach \x in {0,-\distance,...,-\cubex}
        \draw[black] (\x,-\cubey,0) -- (\x,-\cubey-\linelength,\inclination) ;
    \foreach \x in {0,-\distance,...,-\cubex}
        \draw[black] (0,-\cubey,\x) -- (\inclination,-\cubey-\linelength,\x) ;

    \draw[->,-latex] (-\cubex,0,-\cubez) -- (-\cubex,\arrowlength,-\cubez) node[right,xshift=-0.2em,yshift=0.2em]{$u_D$};
    \draw[->,-latex] (-\cubex,0,0) -- (-\cubex,\arrowlength,0);
    \draw[->,-latex] (0,0,0) -- (0,\arrowlength,0);
    \draw[->,-latex] (0,0,-\cubez) -- (0,\arrowlength,-\cubez);

    \draw[] (\shift,0,0) -- (\shift,-\cubey/2,0) node[midway,below,rotate=90]{\smaller $(1-\kappa)L$} -- (\shift,-\cubey,0) node[midway,below,rotate=90]{\smaller $\kappa L$};

    \draw  node[fill=red!40!gray,circle,inner sep=0pt,minimum size=\nodesize pt] at (0,0,0) {};
    \draw  node[fill=red!40!gray,circle,inner sep=0pt,minimum size=\nodesize pt] at (-\cubex,0,0) {};
    \draw  node[fill=red!40!gray,circle,inner sep=0pt,minimum size=\nodesize pt] at (-\cubex,0,-\cubez) {};
    \draw  node[fill=red!40!gray,circle,inner sep=0pt,minimum size=\nodesize pt] at (0,0,-\cubez) {};

    \draw  node[fill=blue!40!gray,circle,inner sep=0pt,minimum size=\nodesize pt] at (0,-\cubey/2,0) {};
    \draw  node[fill=blue!40!gray,circle,inner sep=0pt,minimum size=\nodesize pt] at (-\cubex,-\cubey/2,0) {};
    \draw  node[fill=blue!40!gray,circle,inner sep=0pt,minimum size=\nodesize pt] at (-\cubex,-\cubey/2,-\cubez) {};
    \draw  node[fill=blue!40!gray,circle,inner sep=0pt,minimum size=\nodesize pt] at (0,-\cubey/2,-\cubez) {};

    \draw  node[fill,circle,inner sep=0pt,minimum size=\nodesize pt] at (0,-\cubey,0) {};
    \draw  node[fill,circle,inner sep=0pt,minimum size=\nodesize pt] at (-\cubex,-\cubey,0) {};
    \draw  node[fill,circle,inner sep=0pt,minimum size=\nodesize pt] at (-\cubex,-\cubey,-\cubez) {};
    \draw  node[fill,circle,inner sep=0pt,minimum size=\nodesize pt] at (0,-\cubey,-\cubez) {};
\end{tikzpicture}
        }
        {
         \includegraphics{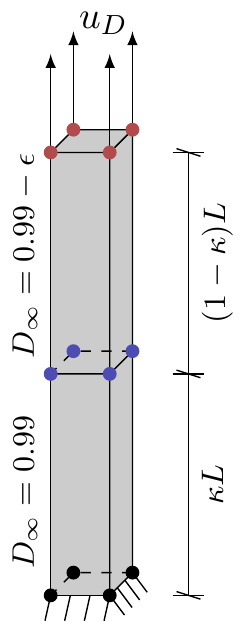}
        }
     \end{minipage}
     \begin{minipage}{0.76\textwidth}
         \centering
        \ifthenelse{\boolean{genfig}}
        {
         \input{figures/tikz/perturbation-restructure-3d-reconvexify}
        }
        {
        \includegraphics{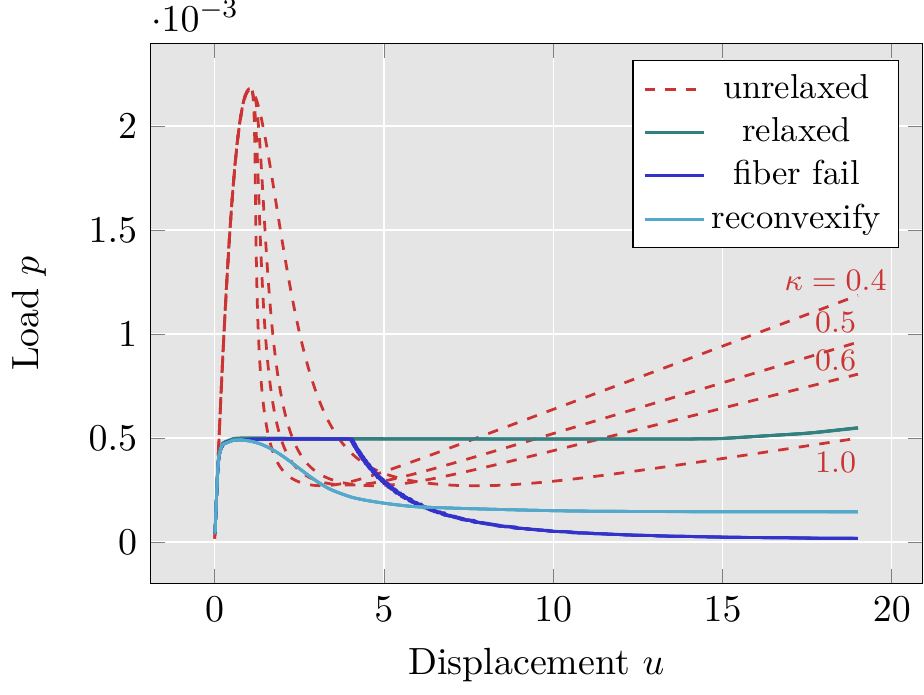}
        }
     \end{minipage}
     \caption{Three-dimensional two element perturbation problem setup on the left-hand side and force-displacement plot on the right-hand side. 
     Plotted are the results for the relaxed formulations \cite{BalOrt:2012:riv}, \cite{KohNeuPetPetBal:2021:easa}, and the proposed approach. Whereas the fiber fail model from \cite{KohNeuPetPetBal:2021:easa} has a rather sudden initialization of strain softening, the proposed model shows a more realistic response.} 
     \label{fig:perturbation3d}
\end{figure}

In order to analyze the mesh independence, the two element material perturbation test is transferred to the three dimensional case by using trilinear continuum elements instead of linear one-dimensional ones. 
On the left-hand side of Figure~\ref{fig:perturbation3d} the discretized boundary value problem is depicted.
Again, the problem consists of two elements, where the upper element is distorted by a small perturbation $\epsilon$. 
However, in contrast to the one-dimensional perturbation test, there are now 12 degrees of freedom in the middle, where eight of them, corresponding to $x_2$ and $x_3$ directed displacements, are prescribed to zero by Dirichlet boundary conditions. 
The nodes with degrees of freedom in the middle of the domain are drawn as blue circles.
At the bottom of the domain, the black drawn nodes, depict the fixed nodes and the top nodes, filled in red, are prescribed in all three components by a Dirichlet boundary condition that increases linearly over the incremental steps. 
The unit sphere integration scheme of choice is the $61\times 2$ symmetric integration scheme reported in \cite{BazOh:1986:eni}. 
From Figure~\ref{fig:perturbation3d} it can be concluded that for a simple boundary value probelm, the integrated response of one-dimensional fibers with the reconvexified model is mesh independent. 
In contrast to the fiber failure approach of \cite{KohNeuPetPetBal:2021:easa}, where a sudden failure of single fibers enables the description of macroscopic strain-softening, the strain-softening mechanism itself is now incorporated already in the fiber description. 
Compared to the fiber failure approach of the aforementioned publication, where a slight mesh-dependency due to the discontinuous response in each fiber was observed, the new approach shows no mesh dependence at all within this simplistic test. 
Furthermore, since the fiber failure is omitted, there is no need to use extremely accurate unit sphere integration schemes and by that, the simulation time is significantly reduced, such that more complex, realistic boundary value problems become more feasible. 
\subsubsection{Plate with a Hole}
\begin{figure}[t]
     \centering
     \begin{subfigure}[b]{0.47\textwidth}
          \centering
        \ifthenelse{\boolean{genfig}}
        {
          \input{figures/platehole/domain_annotated.tex}
        }
        {
          \includegraphics{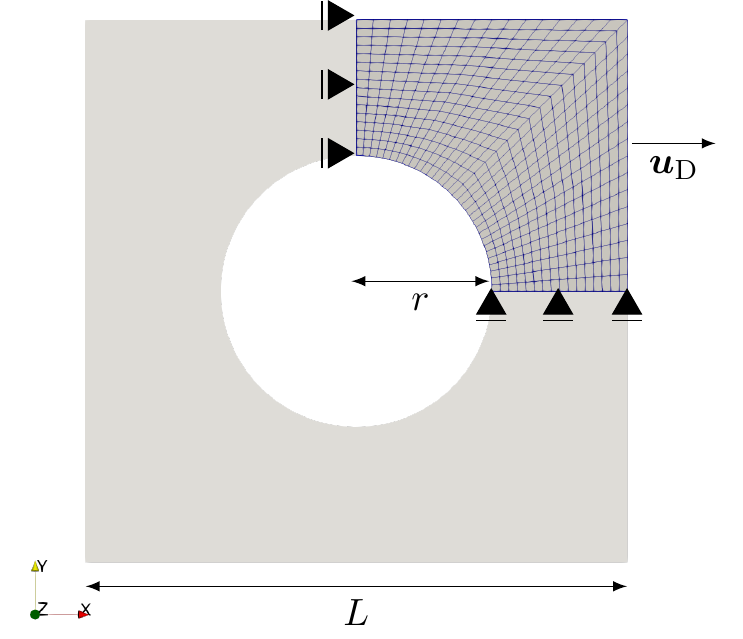}
        }
     \end{subfigure}
     \hfill
     \begin{subfigure}[b]{0.47\textwidth}
          \centering
        \ifthenelse{\boolean{genfig}}
        {
          \tikzsetfigurename{quarterplate_gp}
\begin{tikzpicture}
\begin{axis}[width={0.95\textwidth}, height={0.95\textwidth}, xlabel={Deformation Gradient $F_{11}$}, ylabel={Cauchy Stress $\sigma_{11}$}, xlabel style={font={\normalsize}}, ylabel style={font={\normalsize}}, legend style={font={\small}}, yticklabel style={font={\normalsize}}, xticklabel style={font={\normalsize}}, axis background/.style={fill={white!89.803921568!black}}, x grid style={white}, y grid style={white}, xmajorgrids, ymajorgrids, legend pos={north east}]
    \addplot+[no markers, color={purple!60!gray}, ultra thick]
        table
        {figures/tikz/rawdata/fiberplatehole_8_ele5_gp4_sigma_F.txt};
\end{axis}
\end{tikzpicture}
        }
        {
         \includegraphics{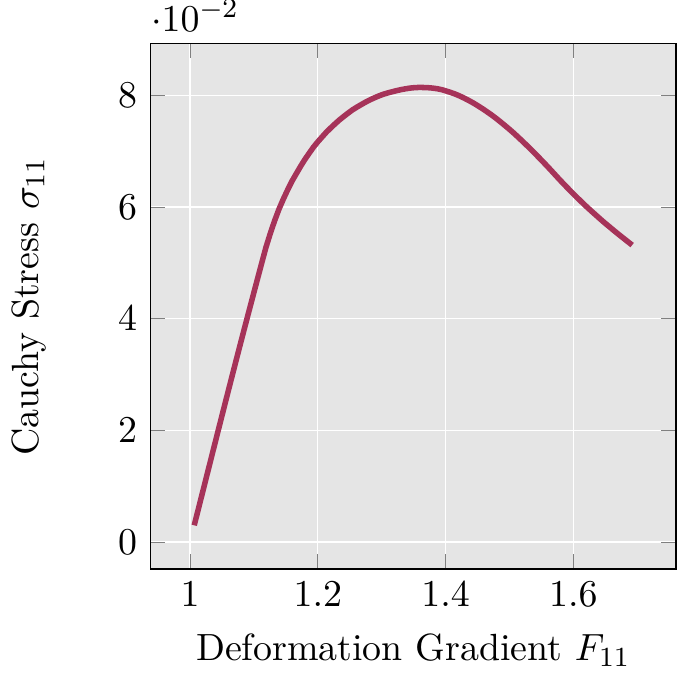}
        }
     \end{subfigure}
     \caption{
     Plate with a hole: (left) boundary value problem 
     of a plate with a hole under plane strain conditions with finest discretization of the upper right part with 512 biquadratic elements. 
     (right) response in a single Gauss point at the top left-hand side of the circular cut out.}
     \label{fig:platehole}
 \end{figure}

\begin{figure}[!h]
     \centering
     \begin{subfigure}[t]{0.47\textwidth}
          \centering
        \ifthenelse{\boolean{genfig}}
        {
          \input{figures/tikz/force_displacement_fiberplatehole_unrelaxed}
        }
        {
          \includegraphics{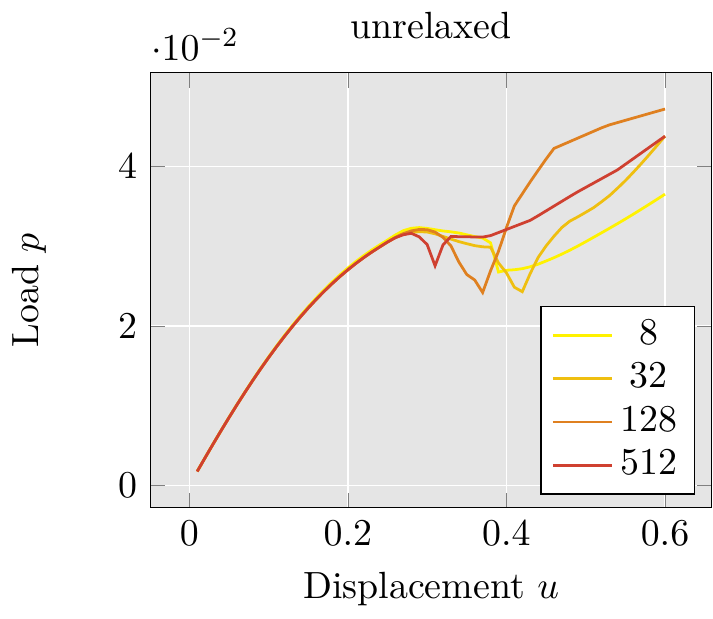}
        }
     \end{subfigure}
     \hfill
     \begin{subfigure}[t]{0.47\textwidth}
          \centering
        \ifthenelse{\boolean{genfig}}
        {
          \input{figures/tikz/force_displacement_fiberplatehole_reconvexify}
        }
        {
         \includegraphics{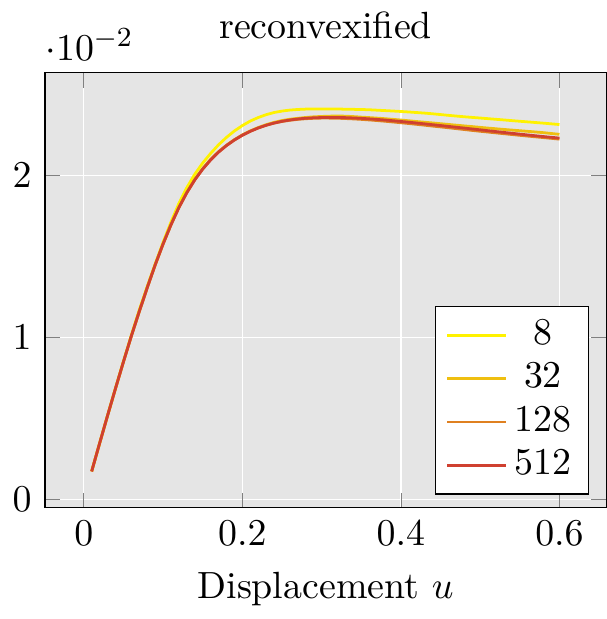}
        }
     \end{subfigure}
     \caption{Force-displacement curve of the plate with a hole problem for the unrelaxed material description (left) and the reconvexified material model (right). 
     Clearly, the unrelaxed model {shows} a significant mesh dependency, whereas the the reconvexified model does not, while still allowing for strain softening.}
     \label{fig:platehole_forcedisplacement}
 \end{figure}

\begin{figure}[h]
      \centering
      \begin{subfigure}[t]{0.3\textwidth}
          \centering
          \includegraphics[width=\textwidth]{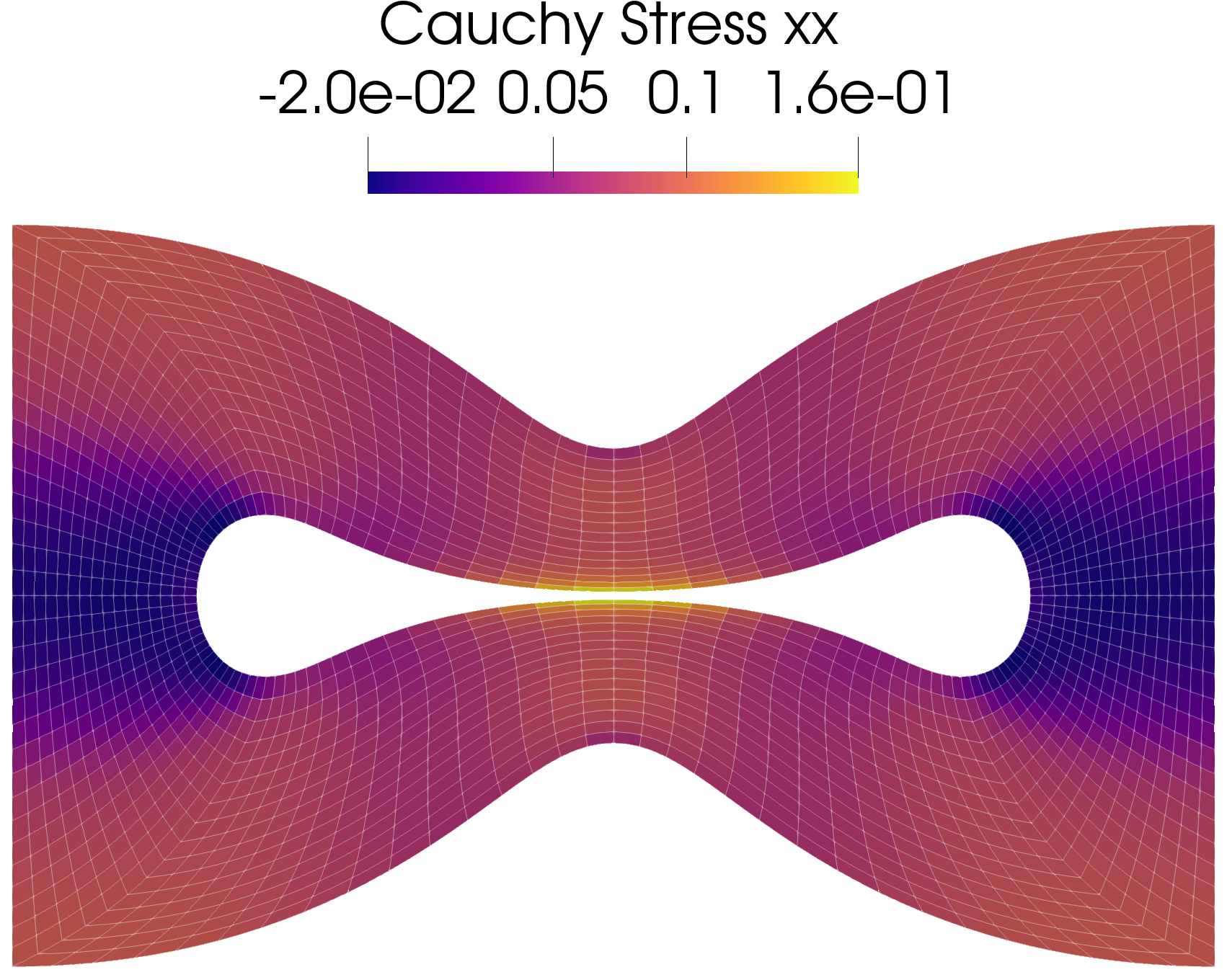}
          \caption{$\sigma_{11}$}
          \label{fig:platehole_sigma_xx}
      \end{subfigure}
      \hfill
      \begin{subfigure}[t]{0.3\textwidth}
          \centering
          \includegraphics[width=\textwidth]{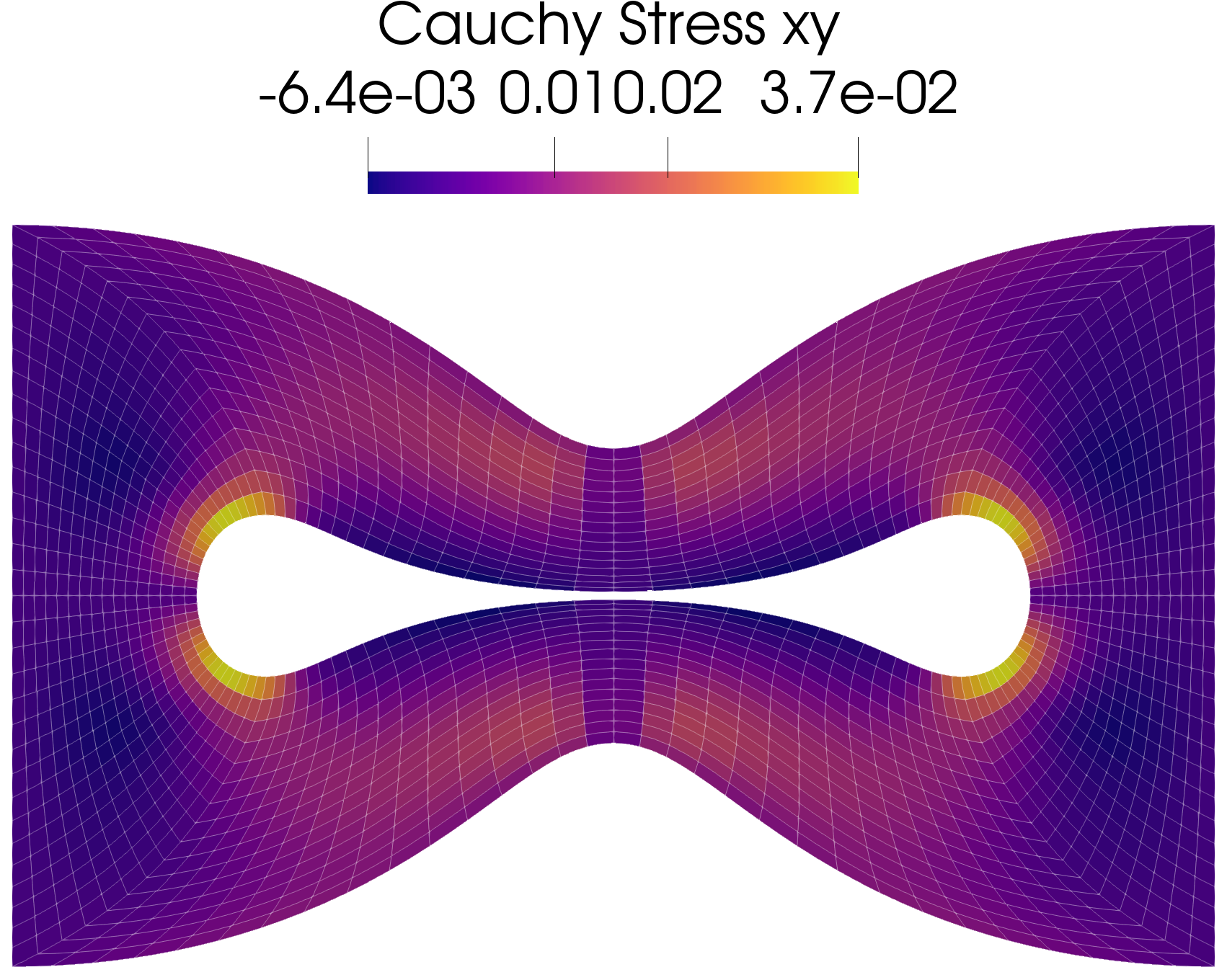}
          \caption{$\sigma_{12}$}
          \label{fig:platehole_sigma_xy}
      \end{subfigure}
      \hfill
      \begin{subfigure}[t]{0.3\textwidth}
          \centering
          \includegraphics[width=\textwidth]{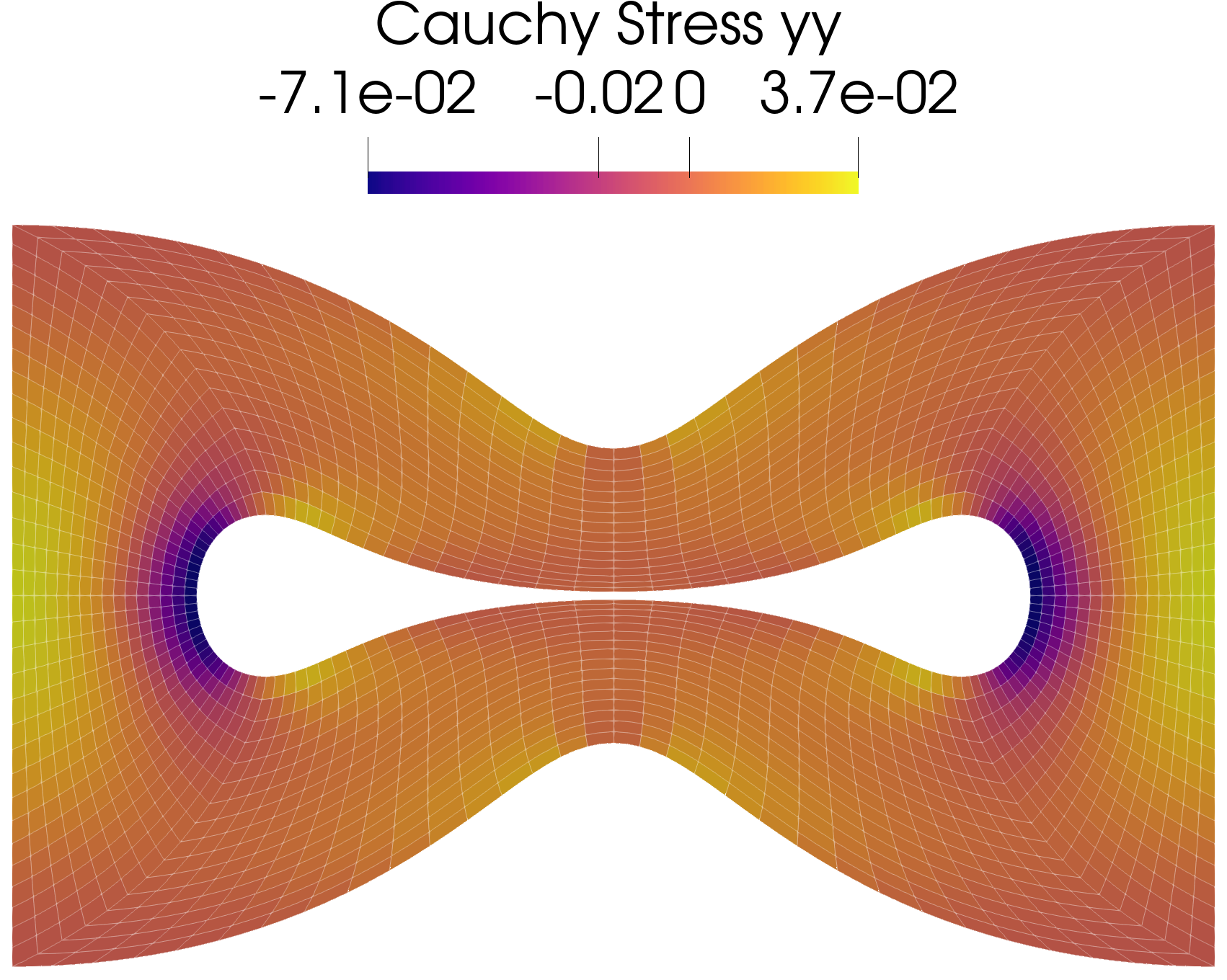}
          \caption{$\sigma_{22}$}
          \label{fig:platehole_sigma_yy}
      \end{subfigure}
         \caption{Deformed final configuration of the domain with contour plots of the element average Cauchy stresses for the finest discretization and the novel reconvexified model.}
         \label{fig:platehole_deformed_cauchy}
 \end{figure}

\begin{figure}[h]
      \centering
      \begin{subfigure}[t]{0.23\textwidth}
          \centering
          \includegraphics[width=\textwidth]{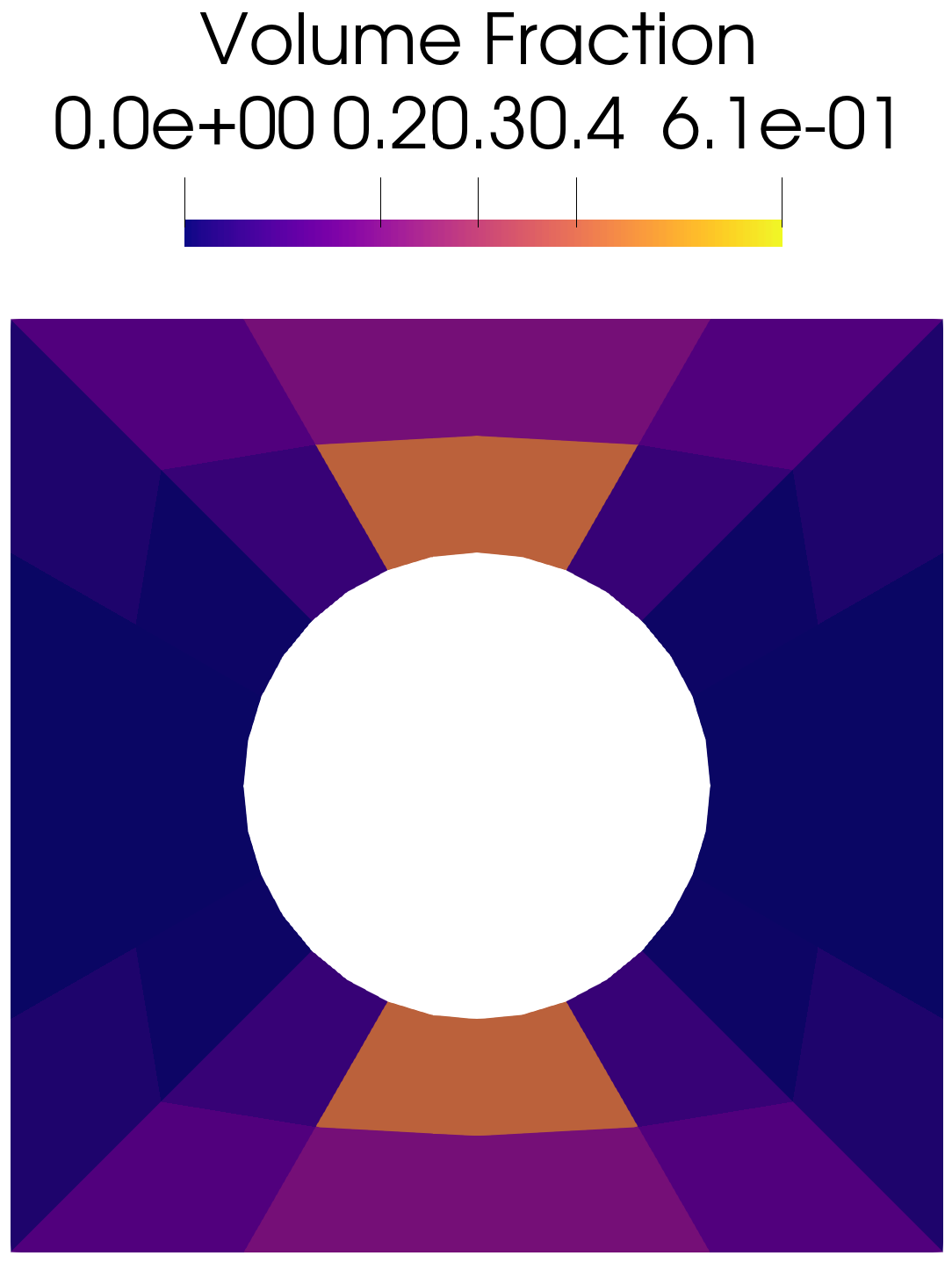}
          \caption{8 elements}
          \label{fig:platehole_xi_8}
      \end{subfigure}
      \hfill
      \begin{subfigure}[t]{0.23\textwidth}
          \centering
          \includegraphics[width=\textwidth]{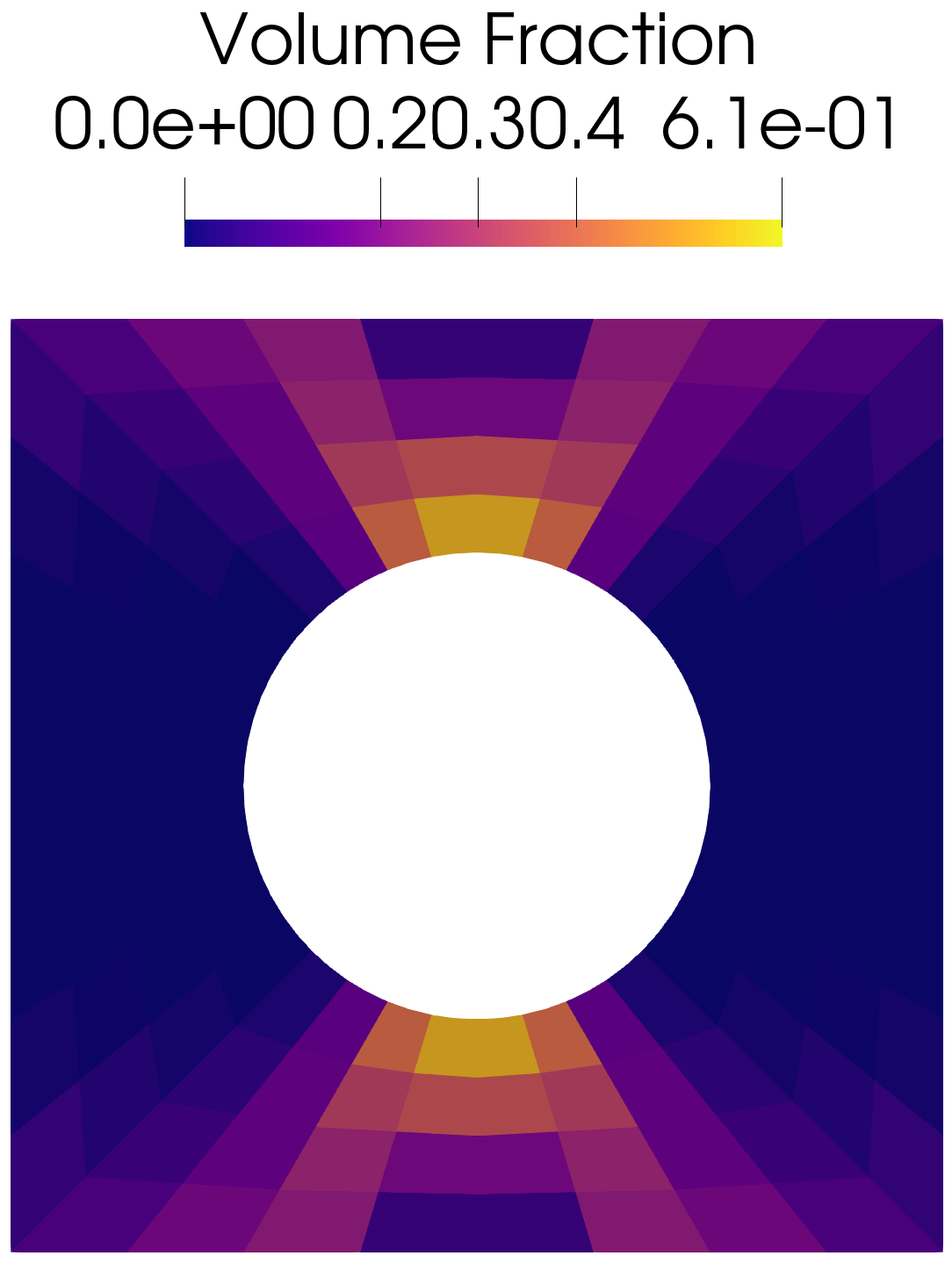}
          \caption{32 elements}
          \label{fig:platehole_xi_32}
      \end{subfigure}
      \hfill
      \begin{subfigure}[t]{0.23\textwidth}
          \centering
          \includegraphics[width=\textwidth]{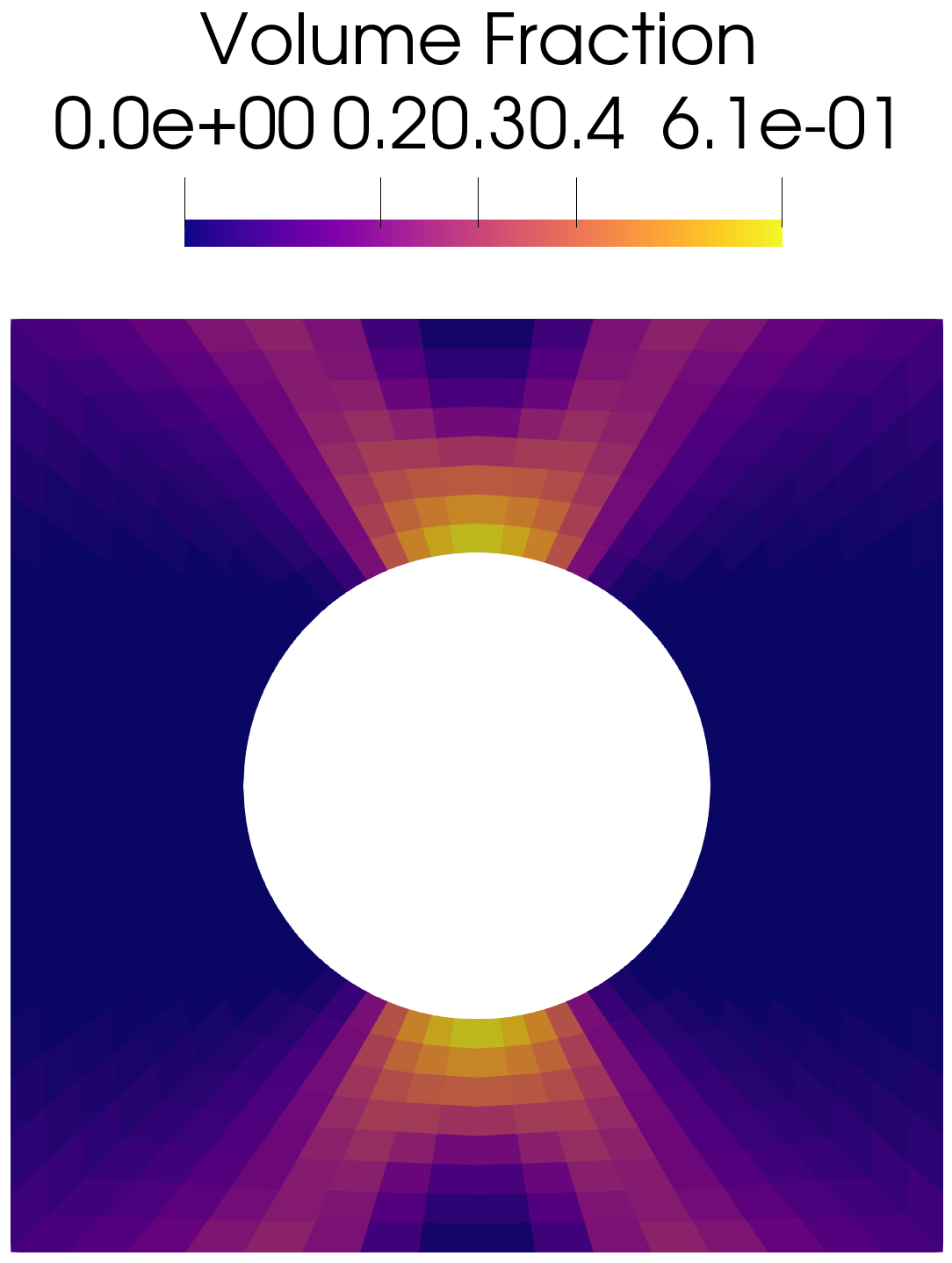}
          \caption{128 elements}
          \label{fig:platehole_xi_128}
      \end{subfigure}
      \hfill
     \begin{subfigure}[t]{0.23\textwidth}
          \centering
          \includegraphics[width=\textwidth]{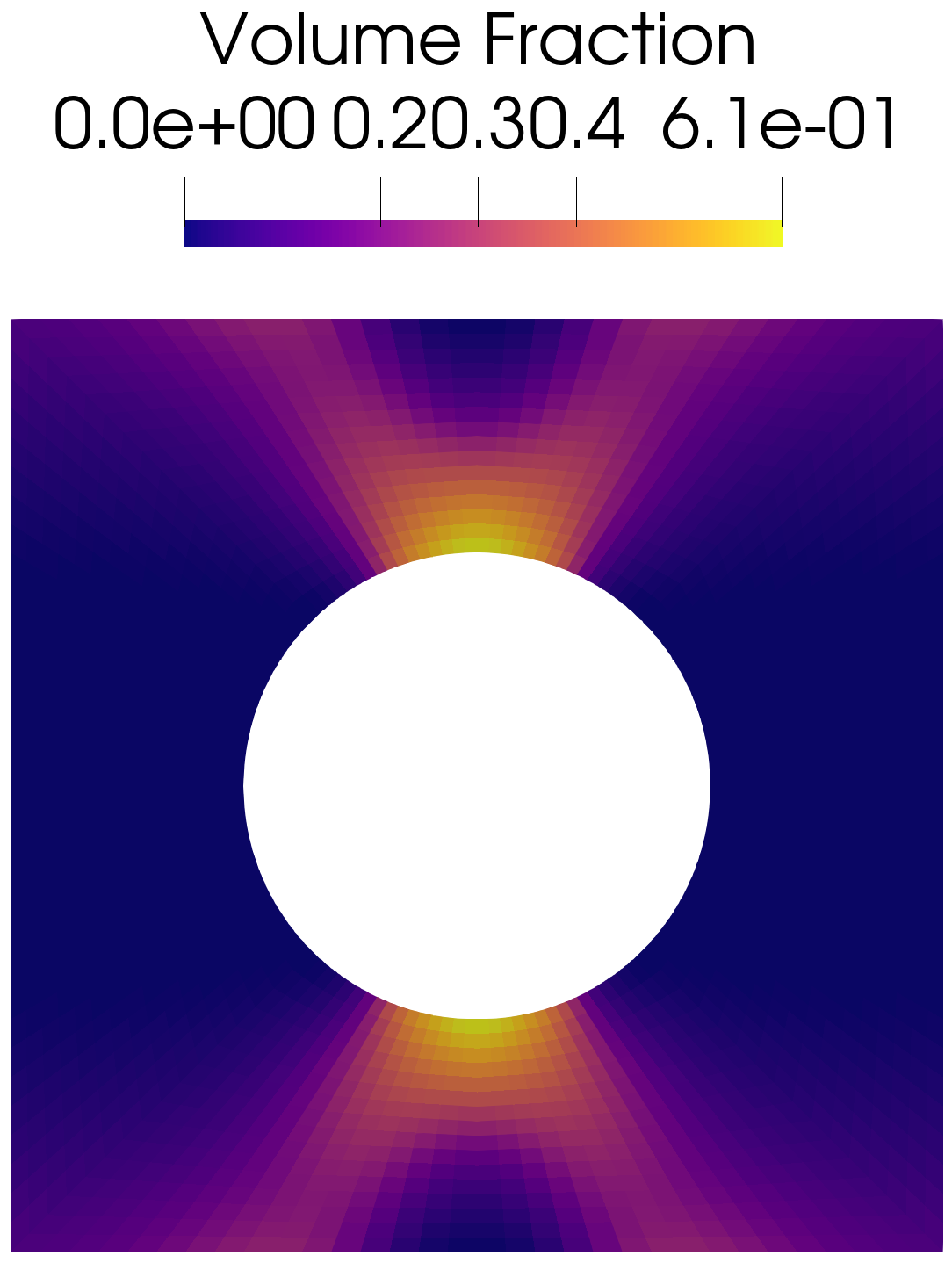}
          \caption{512 elements}
          \label{fig:platehole_xi_512}
      \end{subfigure}
         \caption{
         Contour plots of the volume fraction~$\xi$ (computed as element-wise volume average of the integrals over the unit sphere) for four different discretizations.}
         \label{fig:platehole_deformed_xi}
 \end{figure}

In this subsection the novel reconvexified model is analyzed in the classical benchmark problem of a plate with a hole. 
The boundary value problem is depicted in Figure~\ref{fig:platehole}. 
Due to the symmetry of the problem, only the upper right part needs to be discretized. 
To impose the symmetry in the finite element computation, the degrees of freedom orthogonal to the symmetry line need to be constrained to zero. 
Further, at the right-hand side boundary, an incrementally increasing Dirichlet boundary condition is prescribed for the $x_1$ direction. 
Therefore, the body is allowed to contract transversely to the prescribed boundary condition. 
A neo-Hookean effective energy density~$\psi^0_{\text{NH}}$ is used with the Lamé parameters $\lambda=0.5$ and $\mu=1.0$ and the damage parameters $D_0=0.2,D_\infty=0.9$. 
For the generation of the structured biquadratic meshes, Gmsh~\cite{GeuRem:2009:gtf} is used. 
The plane strain assumption is applied, such that biquadritic quadrilateral elements can be used. 
The resulting non-linear structural problem is solved by a Newton scheme with an Armijo-Goldstein criterion-based linesearch, see e.g. \cite[Sec.  9.2.5]{Bar:2015:nmn}. 
For this boundary value problem, an integration scheme of \cite{SloWom:2004:esp} with 225 integration points on the unit sphere was used, since this type of integration scheme outperformed other used schemes as reported in \cite{KohNeuPetPetBal:2021:easa}. 
Figure~\ref{fig:platehole_forcedisplacement} shows the force-displacement curves of the unrelaxed and reconvexified model. 
On the left-hand side, the response of the unrelaxed model can be seen, which shows a similar behavior as the small-strain unrelaxed model in~\cite[Figure 13a]{GurMie:2011:edm}. 
The right-hand side of Figure~\ref{fig:platehole_forcedisplacement} visualizes the force-displacement curve of the reconvexified model. 
Compared to~\cite[Figure 13b]{GurMie:2011:edm}, i.e. small strain relaxed formulation, the proposed reconvexified model shows a response that includes strain softening. 
In Figure~\ref{fig:platehole_deformed_cauchy}, the finest discretization is visualized in its deformed configuration with a contour plot of the different element average Cauchy stresses. 
From the figure it can be seen that no strain localization occurs, which can be seen as well in the element average volume fraction $\xi$ distribution in Figure~\ref{fig:platehole_deformed_xi}. 
The element average volume fraction is obtained by integrating the microsphere $\xi$ values over the unit sphere with the same integration scheme which was used to obtain the three-dimensional material response. 
This benchmark problem shows that the proposed model is indeed able to describe strain softening while still showing mesh-independent results.
\subsubsection{Inhomogeneous Shear Test}
\begin{figure}[h]
     \centering
     \hspace{-0cm}
     \begin{minipage}{0.48\textwidth}
         \hspace{0.5cm}
         \includegraphics[width=0.9\textwidth]{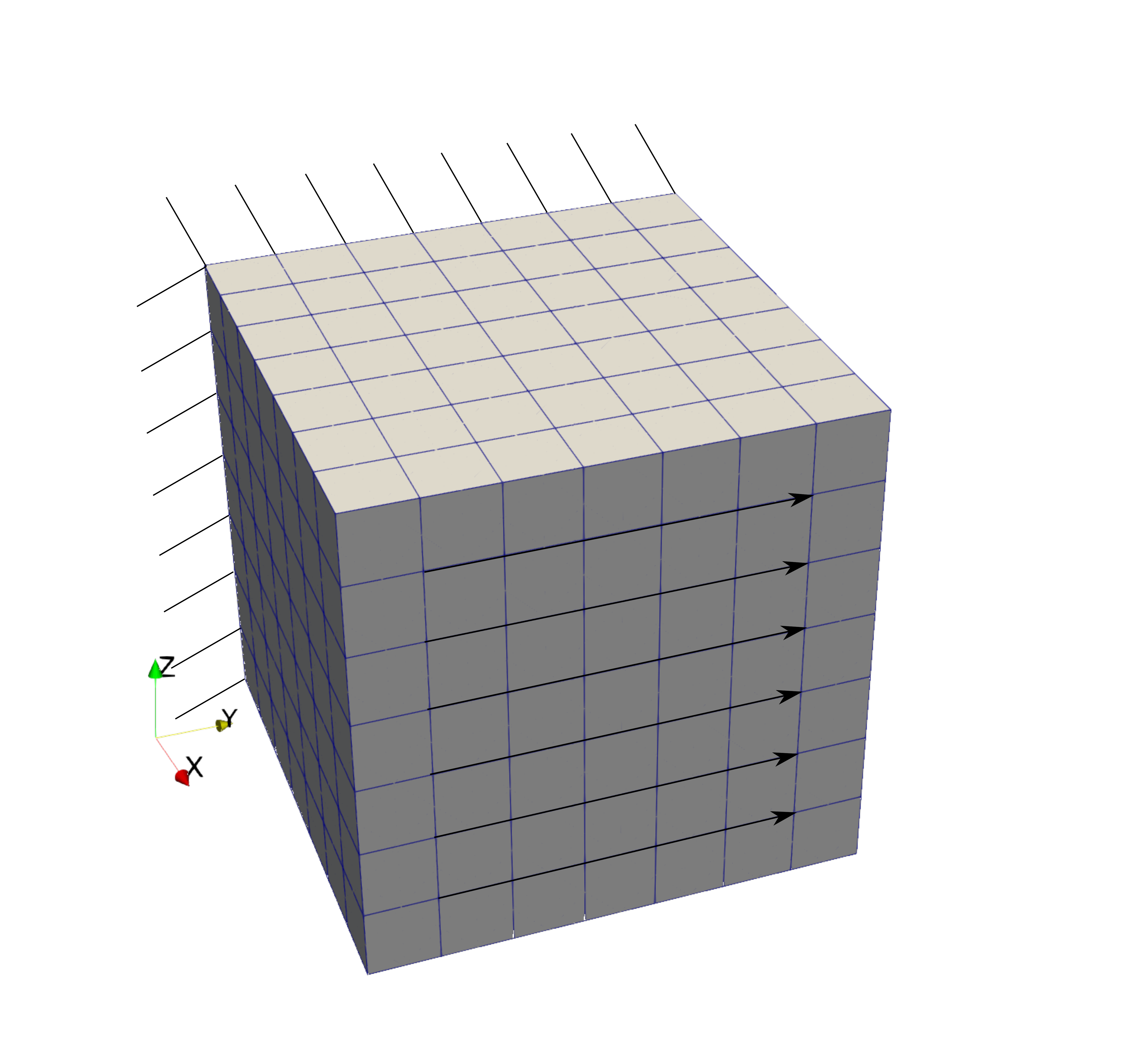}
     \end{minipage}
     \begin{minipage}{0.48\textwidth}
         \centering
         \includegraphics[width=0.9\textwidth]{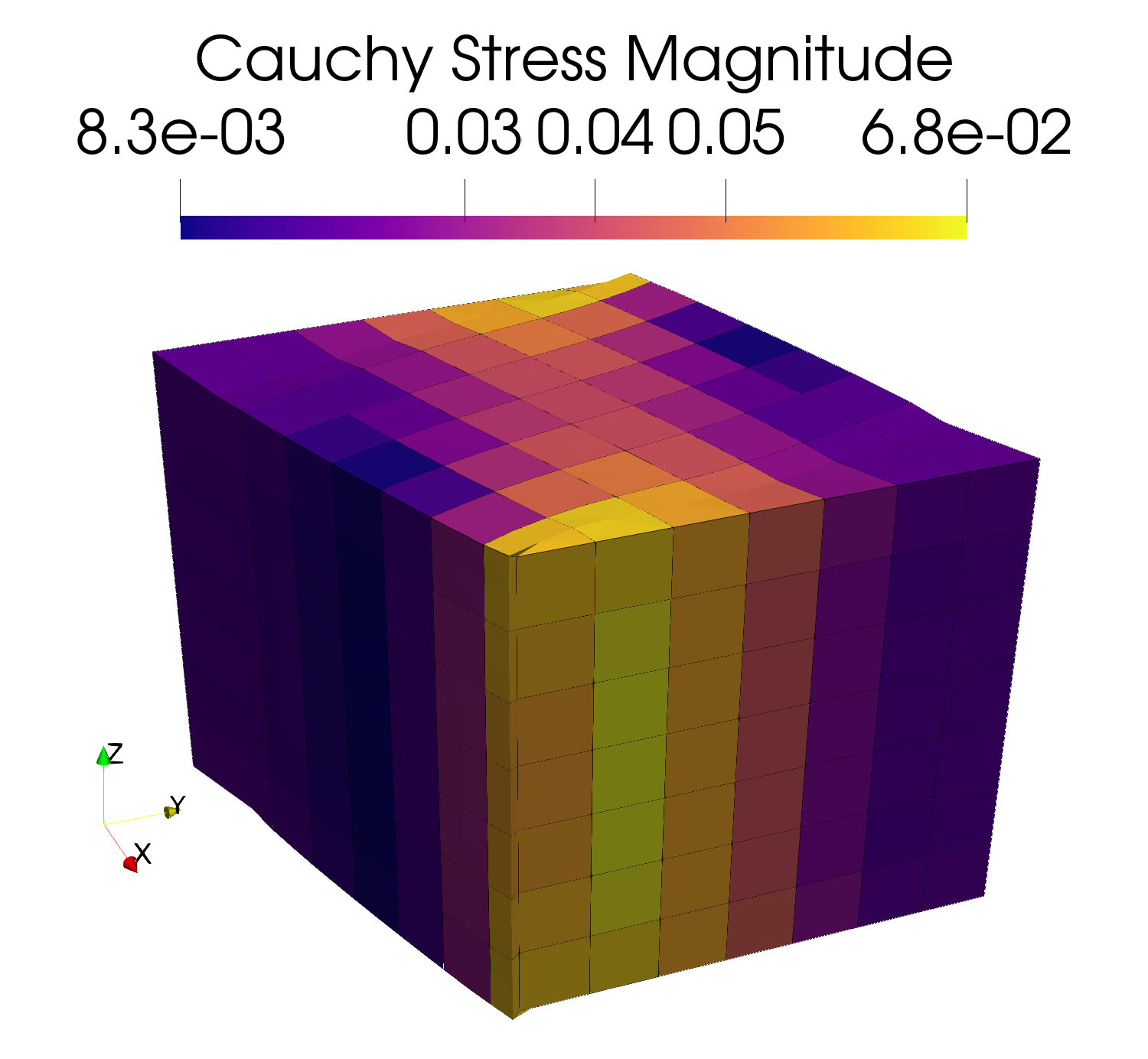}
     \end{minipage}
     \caption{
     Boundary value problem of inhomogeneous shear test (left) and contour plot of the element-average Cauchy stress magnitude in the deformed configuration.}
     \label{fig:shear-stress-defo}
 \end{figure}

A fourth mesh independence test is carried out which analyzes the behavior in a fully three-dimensional setting. 
In order to do so, a unit cube is fixed, i.e. homogeneous Dirichlet boundary conditions are prescribed at the $x=0$ plane and an incremental, linearly increasing non-zero Dirichlet boundary condition at the $x=1$ plane is applied. 
Here, the Dirichlet boundary condition is non-zero in the second component, i.e. $y$ component, and zero in the $x$ and $z$ component. 
In Figure~\ref{fig:shear-stress-defo}, the boundary value problem is visualized. 
Here, schematically, linear elements are depicted. 
However, within the simulation, Serendipity elements are used. 
In this problem, the Yeoh type effective strain energy density~$\psi^0_{\text{Yeoh}}$ is used with the material parameters $D_0=0.15$, $D_\infty=0.99$, $c_1=1.0$, $c_2=0.2$, and $c_3=2.0$. 
Again, the integration scheme of \cite{SloWom:2004:esp} with 225 integration points is used.
On the right-hand side of Figure~\ref{fig:shear-stress-defo}, the deformed configuration of the domain with a contour plot of the cell average Cauchy stress magnitude is depicted. 
 \begin{figure}[h]
     \begin{minipage}{0.49\textwidth}
         \centering
        \ifthenelse{\boolean{genfig}}
        {
         \input{figures/tikz/shear-unrelaxed-yeoh_0.5}
        }
        {
         \includegraphics{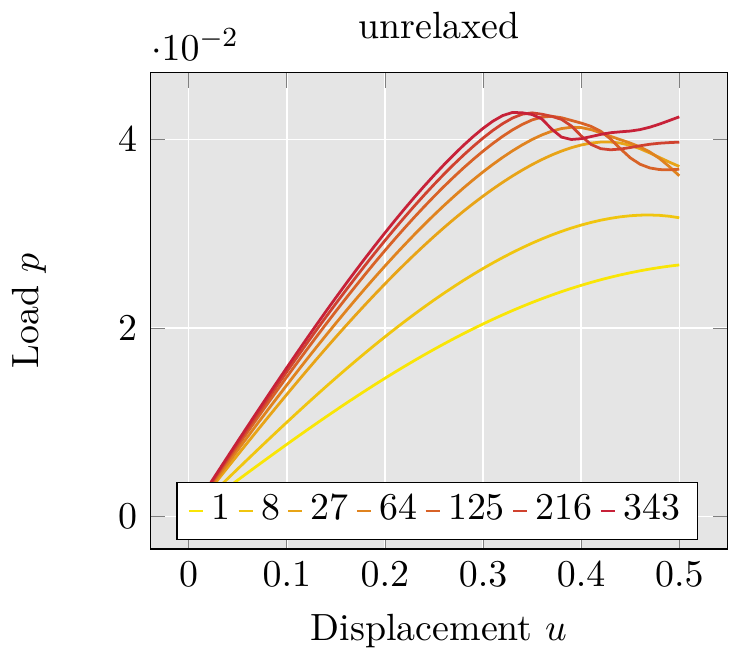}
        }
     \end{minipage}
     \begin{minipage}{0.49\textwidth}
         \centering
        \ifthenelse{\boolean{genfig}}
        {
         \input{figures/tikz/shear-reconvexified-yeoh_0.5}
        }
        {
         \includegraphics{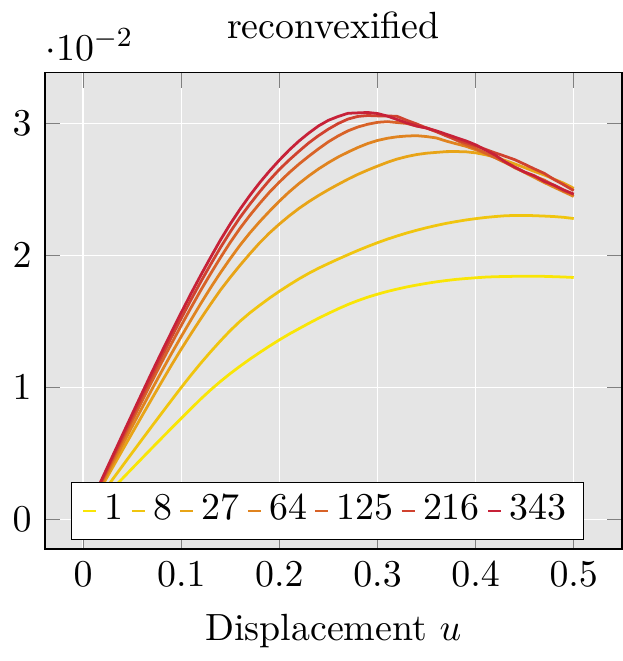}
        }
     \end{minipage}
     \caption{Force--displacement diagrams for different amounts of serendipity elements used in the finite element discretization of the inhomogeneous shear problem.
     The left-hand side shows the unrelaxed response and the right-hand side visualizes the response of the reconvexified model.}
     \label{fig:shear-displacement-force}
 \end{figure}
Figure~\ref{fig:shear-displacement-force} shows the force--displacement curve for different discretizations ranging from one element in each axis of the three dimensional space to seven elements in each axis. 
On the left-hand side of the figure, the unrelaxed response is plotted, whereas the right-hand side depicts the response of the reconvexified model. 
From this figure it can be concluded that the overall behavior of the reconvexified model is converging. 
As in the previous two-dimensional example, the results for this three-dimensional problem are similar. 
In contrast to previous relaxed formulations, the proposed model allows for the robust description of strain-softening while ensuring mesh-independent results. 

\clearpage
\section{Conclusion
\label{sec:conclusion}}
In this contribution, a novel reconvexified model was proposed which extends the relaxation approaches from \cite{BalOrt:2012:riv}, \cite{GurMie:2011:edm}, and \cite{KohNeuPetPetBal:2021:easa}. 
The main idea was to consider an updated convexification in each incremental step rather than fixating the convex hull as soon as {convexity is lost}.
The gradient Young measure was modeled such that on the one hand, the volume fraction~$\xi$ remains irreversible and on the other hand, constant internal variables for the strongly and weakly damaged phase{s} are obtained. 
Because of this, the volume fraction~$\xi$ can be well interpreted.
The constraint was enforced only for the weakly damaged phase, while for the strongly damaged phase, the internal variable turned out to automatically stay at the same energetic level. 
This was due to the fact, that the strongly damaged state describes always the border of the non-convex regime which corresponds to the end of the damaging regime.
However, the deformation gradient in the strongly damaged phase could 
change in subsequent incremental steps, leading to a non-linear irreversible behavior of~$\xi$ as well as a successive decreasing slope of the convex hull. 
This fact allowed the model to describe strain softening. 
In order to show that a realistic strain softening response can indeed be described, the proposed model was adjusted by hand to experimental data performed on tissue samples of an abdominal aortic aneurysm and a reasonable agreement with the experiments could be shown.
In several numerical examples including one-, two- and three-dimensional inhomogeneous problems, the reconvexification approach was analyzed. 
There, it could be shown that significant strain softening can be described robustly while keeping the finite element simulations mesh independent. 
In contrast to the previous relaxed formulation in~\citet{KohNeuPetPetBal:2021:easa}, here, a robust approach was proposed which does not suffer from the problems reported in~\cite{KohNeuPetPetBal:2021:easa}. 
These included a discontinuous response in the single fiber directions leading to problems regarding the calculation of suitable tangent moduli and requiring significant effort on the microsphere integration. 
In the present approach, these issues do not appear and enabled the calculation of relevant boundary value problems. 
Future topics related to this contribution are the mathematical analysis of the model as well as further numerical studies about the model behavior. 
Furthermore, here, the three-dimensional material response was obtained from microsphere integration. 
However, the ideas proposed here can in principle also be directly applied in an, admittedly much more expensive, two- or three-dimensional convexification framework. 
In addition to that, the adaptive convexification procedure of \cite{KohNeuPetPetBal:2021:easa} should be applied to the reconvexification procedure as well. 
Then the adaptive polynomial approach only need to be applied once and afterwards the convexification grid nodes can be shifted accordingly to the change of shape of the incremental stress potential in the subsequent steps.

\section*{Acknowledgement}

We acknowledge the Deutsche Forschungsgemeinschaft (DFG) for funding within the Priority Program 2256 ("Variational Methods for Predicting Complex Phenomena in Engineering Structures and Materials"), project ID 441154176, reference ID BA2823/17-1. 
Furthermore, the free and open source software community, especially of the Julia programming language and Ferrite.jl, as well as scientific discussions with Dennis Ogiermann, Timo Neumeier, Malte A. Peter, and Daniel Peterseim are greatly acknowledged.

\bibliographystyle{stylefiles/references_style}
\bibliography{references}

\end{document}